\documentclass[reprint,amsmath,amssymb,prb,aps]{revtex4-1}

\bibpunct{[}{]}{,}{n}{}{,}

\usepackage[utf8x]{inputenc}

\usepackage{siunitx}
\usepackage{graphicx}
\usepackage{hyperref}
\usepackage[expansion=false]{microtype}

\newcommand{\mytitle}{%
  Structural origins of the boson peak in metals:\\ From high-entropy
  alloys to metallic glasses
}
\hypersetup{
  unicode, final,
  pdfborder={0 0 0},
  pdftitle={\mytitle},
  pdfauthor={Tobias Brink, Leonie Koch, and Karsten Albe},
  colorlinks=true,
  urlcolor=blue,
  linkcolor=blue,
  citecolor=blue
}

\begin{document}
\frenchspacing
\raggedbottom

\title{\mytitle}

\author{Tobias Brink}
\email{brink@mm.tu-darmstadt.de}
\affiliation{Fachgebiet Materialmodellierung, Institut f{\"u}r
  Materialwissenschaft, TU Darmstadt,
  Jovanka-Bontschits-Stra\ss{}e~2, D-64287 Darmstadt, Germany}

\author{Leonie Koch}
\affiliation{Fachgebiet Materialmodellierung, Institut f{\"u}r
  Materialwissenschaft, TU Darmstadt,
  Jovanka-Bontschits-Stra\ss{}e~2, D-64287 Darmstadt, Germany}

\author{Karsten Albe}
\affiliation{Fachgebiet Materialmodellierung, Institut f{\"u}r
  Materialwissenschaft, TU Darmstadt,
  Jovanka-Bontschits-Stra\ss{}e~2, D-64287 Darmstadt, Germany}

\date{December 14th, 2016}

\begin{abstract}
  The boson peak appears in all amorphous solids and is an excess of
  vibrational states at low frequencies compared to the phonon
  spectrum of the corresponding crystal. Until recently, the consensus
  was that it originated from ``defects'' in the glass. The nature of
  these defects is still under discussion, but the picture of regions
  with locally disturbed short-range order and/or decreased elastic
  constants has gained some traction.  Recently, a different theory
  was proposed: The boson peak was attributed to the first van Hove
  singularity of crystal lattices which is only smeared out by the
  disorder. This new viewpoint assumes that the van Hove singularity
  is simply shifted by the decreased density of the amorphous state
  and is therefore not a glass-specific anomaly. In order to resolve
  this issue, we use computer models of a four-component alloy,
  alternatively with chemical disorder (high-entropy alloy),
  structural disorder, and reduced density.  Comparison to a reference
  glass of the same composition reveals that the boson peak consists
  of additional vibrational modes which can be induced solely by
  structural disorder.  While chemical disorder introduces
  fluctuations of the elastic constants, we find that those do not lead
  to sufficient local softening to induce these modes.  A boson peak
  due to a reduction of density could be excluded for the present
  metallic system.\\[-1.5\baselineskip]
  \begin{center}
    \makeatletter
    \rule{\frontmatter@abstractwidth}{0.4pt}
    \makeatother
  \end{center}
  \noindent
  \footnotesize
  Published in:\\
    \href{https://doi.org/10.1103/PhysRevB.94.224203}
         {T.~Brink \textit{et al.},
          Phys.\ Rev.\ B \textbf{94}, 224203 (2016)}
    \hfill
    DOI: \href{https://doi.org/10.1103/PhysRevB.94.224203}
              {10.1103/PhysRevB.94.224203}\\%
    \copyright{} 2016 American Physical Society.

\end{abstract}

\maketitle

\section{Introduction}

Amorphous materials show an excess contribution in the terahertz
region of the vibrational density of states (VDOS) as compared to the
corresponding crystal or the Debye model. The origin of this so-called
boson peak is still a matter of ongoing discussion.  Early
explanations include scattering of phonons at density fluctuations
\cite{Elliott1992} and the soft potential model \cite{Karpov1983,
  Laird1991, Schober1991, Gil1993, Schober1996}. The latter assumes an
interaction of acoustic phonons with quasi-localized modes arising
from ``defects'' in the disordered structure \cite{Shintani2008,
  Schober2011, Schober2014}.  It was proposed that these defects are
loosely packed atoms \cite{Schober1996, Li2006, Guerdane2008,
  Jakse2012, Sheng2012} or resemble interstitialcies
\cite{Granato1996, Vasiliev2009}.  An alternative explanation, where
the boson peak is considered to be a precursor of a dynamical
instability, was brought forward by Grigera \textit{et al.}
\cite{Grigera2003}, while Schirmacher and colleagues developed a
theory where fluctuating force or elastic constants give rise to the
additional modes that make up the boson peak \cite{Schirmacher1993,
  Schirmacher1998, Schirmacher2007, Schirmacher2008, Marruzzo2013,
  Schirmacher2015}.  This viewpoint is supported by studies that
connect the boson peak to ``soft spots,'' i.e., regions of reduced
short-range order, stiffness, and mechanical strength
\cite{Tanguy2010, Sheng2012, Derlet2012, Ding2014b}.  For
silicon-based glasses, similar findings exist \cite{Leonforte2006,
  Fusco2010} and a relation between the bending rigidity of the
silicon tetrahedra, the shear modulus, and the boson peak has been
proposed \cite{Beltukov2016}.

In contrast to these theories, which describe the boson peak as
additional modes present only in glasses, Taraskin \textit{et al.}\
proposed that the boson peak is related to the first van Hove
singularity of the crystal, which is shifted to lower frequencies by
fluctuating force constants \cite{Taraskin2001}.  More recently, work
by Chumakov and colleagues received attention \cite{Chumakov2011,
  Chumakov2014, Chumakov2015, Chumakov2016, Baldi2016}: Measurements
on oxide glasses suggest that the boson peak is simply a van Hove
singularity that is shifted because of the lower volume density of the
glass compared to the crystal. These authors criticize alternative
models like Schirmacher's as they do not explicitly treat the
\mbox{(pseudo-)}Brillouin zone and therefore do not include the
necessary van Hove singularities \cite{Chumakov2016}.  Furthermore,
the argument is that disordered systems are always less dense than
ordered ones. That leads to a lower transversal sound velocity, to a
smaller size of the \mbox{(pseudo-)}Brillouin zone, and in consequence
to a shift to lower frequencies, the result of which was interpreted
as a boson peak \cite{Chumakov2016}. This would equally apply to soft
spots, which are also characterized by a lowered density.

A material class carrying all the ingredients of those models arguing
for a boson peak induced by fluctuations in the elastic medium, while
still being crystalline and similar in density to the crystalline
forms of the constituent elements, are high-entropy alloys
(HEAs). HEAs are metallic alloys consisting of at least four elements
with equal or near equal molar fractions that are considered to be
thermodynamically stable phases because of their large configurational
entropy \cite{Yeh2013, Tsai2014, Miracle2014}.  Single-phase HEAs are
random solid solutions and as such---from the point of view of lattice
dynamics---they can be thought of as a simple crystal lattice with
randomly distributed force constants.  Therefore, they are a suitable
model to discern between the different proposed origins of the boson
peak.  Experimentally, though, HEAs have been mostly studied with
respect to their mechanical and thermodynamic properties, but not
their vibrational properties.  We use molecular dynamics (MD)
simulations on a CuNiCoFe alloy, since MD provides direct access to
vibrational properties and furthermore allows for additional
modifications of the HEA model: Elastic scaling of the volume of the
crystalline matrix, as well as the gradual introduction of disorder
are possible. For the introduction of disorder, we follow the ideas of
interstitialcy theory which proposes that the melting of metals is
comparable to the introduction of increasing numbers of interstitial
defects that remain discernible even in the liquid and glassy states
\cite{Khonik2015}. Using this tunable model, and by comparison with a
metallic glass of the same composition---quenched from the melt at
high cooling rates---, we can identify the origin of the boson peak.

\section{Simulation methods and analysis}

\subsection{Preparation of CuNiCoFe high-entropy alloys and glasses}

We use MD simulations on a CuNiCoFe alloy consisting of
$N = \num{102816}$ atoms with an embedded atom method potential by
Zhou \textit{et al.}  \cite{Zhou2004}.  This provides a realistic
description of the vibrational properties but does not contain
electronic or magnetic effects, which we can safely ignore in
discussing the boson peak. All simulations were performed with
\textsc{lammps} \cite{Plimpton1995}. In order to generate the initial
HEA structure, we used a hybrid Monte Carlo/MD method in the
variance-constrained semi-grand-canonical ensemble \cite{Sadigh2012},
which provides a significant reduction of computation time needed to
reach the equilibrium chemical order.  Here, we performed $N/4$ Monte
Carlo trial moves in between every 20 MD steps at \SI{800}{K}.  The
constraint on the variation of the elemental concentrations allows one
to study phase formation---even in miscibility gaps---and
simultaneously maintain an equimolar composition \cite{Sadigh2012}.
Since element concentrations depend on the excess chemical potentials
$\Delta \mu_i$, the simulation needs a set of chemical potential
differences $\{\Delta \mu_i \}$ as input. If a set $\{\Delta \mu_i \}$
exists, such that the system is still miscible in a random solid
solution, the resulting sample is a good model for a single-phase
HEA. To probe for the chemical potential differences, we disabled the
variance constraint and employed Gaussian processes \cite{Wilson2014},
a global optimization algorithm: We used $\{\Delta \mu_i \}$ as the
parameters and minimized the deviation from equimolar concentrations
obtained by running simulations in the semi-grand-canonical
ensemble. This works because without the variance constraint an
immiscible system will reduce the concentration of one or more
elements, while a miscible system will stay equimolar.  We found a
solution in which the system is in fact miscible: The final chemical
potential differences relative to Cu are
$\Delta \mu_\text{Cu--Ni} = \SI{0.9}{eV}$,
$\Delta \mu_\text{Cu--Co} =\SI{0.85}{eV}$, and
$\Delta \mu_\text{Cu--Fe} = \SI{0.7}{eV}$.

We performed this investigation at \SI{800}{K} because this makes it
easier to find a miscible alloy: The $TS$-term due to chemical
disorder dominates the thermodynamics of the system. This is a
reasonable facsimile of the experimental synthesis, which often starts
from the melt.  If this alloy, and HEAs in general, are still miscible
at lower temperatures---or if they are simply kinetically trapped
during quenching---is an important unsolved question which lies beyond
the scope of this work.

In addition to the HEA sample, we also prepared a glass sample of the
same composition by quenching from the melt at \SI{2000}{K} to
\SI{30}{K} with a cooling rate of \SI{e13}{K/s} (the minimum rate at
which crystallization is still kinetically suppressed).

\subsection{Calculation of the vibrational density of states}

We measured the VDOS by recording the velocity auto-correlation
function during an MD run, the Fourier transform of which is the
density of states $g(\nu)$ \cite{Dickey1969}.  All systems were first
equilibrated for \SI{0.5}{ns} at \SI{30}{K} and ambient pressure; the
velocity auto-correlation function was subsequently measured at
\SI{30}{K} in the micro-canonical ensemble. We use the average of 100
auto-correlation functions to reduce the noise.  To differentiate the
contributions of different structures in the system, we also present
data from partial VDOS calculations. Here, we consider only the
partial velocity auto-correlation function of the atoms of interest to
calculate the VDOS.  While one could argue that phonons are
delocalized over the whole system, computer simulations using similar
schemes \cite{Jakse2012} evidence the effectiveness of this method in
the case of the boson peak: Its modes seem to be localized enough to
be able to separate them out by spatial sampling of the VDOS.

Thermodynamic properties for the low-temperature regime---such as the
internal energy $U$ and the heat capacity $C$---were derived from the
VDOS via the harmonic approximation \cite{Pathria1996}
\begin{align}
  \label{eq:harmonic-approximation-U}
  U &= \frac{3}{2}N\! \int_0^\infty h\nu
       \coth\! \left( \frac{1}{2} \frac{h \nu}{k_\mathrm{B} T} \right)
       g(\nu) \mathrm{d}\nu \quad \text{and} \\
  \label{eq:harmonic-approximation-C}
  C &= 3Nk_\mathrm{B} \int_0^\infty
       \left( \frac{h \nu}{2 k_\mathrm{B} T} \right)^2
       \sinh^{-2}\! \left( \frac{h \nu}{2 k_\mathrm{B} T} \right)
       g(\nu) \mathrm{d}\nu.
\end{align}
The Debye temperature $\Theta$ and frequency $\nu_\mathrm{D}$ can then
be obtained from the zero point energy via \cite{Pathria1996}
\begin{equation}
  \label{eq:zero-point-energy}
  U_0 = \frac{9}{8} N k_\mathrm{B} \Theta = \frac{9}{8} N h \nu_\mathrm{D}.
\end{equation}
Given the condition of $\int_0^{\nu_\mathrm{D}} g_\mathrm{D}(\nu)
\mathrm{d}\nu = 1$, one can obtain the complete VDOS in the Debye
model from $\nu_\mathrm{D}$.

\subsection{Calculation of the shear modulus}

In order to probe the local stiffness of our samples, we calculated
the \SI{0}{K} stiffness tensor for every atom.  This was done using
molecular statics simulations in which we subsequently impose the
stress components $\sigma_{xx}$, $\sigma_{yy}$, $\sigma_{zz}$,
$\sigma_{xy}$, $\sigma_{xz}$, and $\sigma_{yz}$.  For this, the
currently active component $\sigma_i$ was set to \SI{50}{MPa} and all
other components $\sigma_{j\neq i}$ were set to zero.  After static
minimization, we calculate the resulting per-atom strain tensors using
the atomic strain calculation in \textsc{ovito} \cite{Shimizu2007,
  Stukowski2010}. Using Hooke's law in tensor form (Voigt notation),
\begin{equation}
  s_{ij} = \frac{\varepsilon_i}{\sigma_j},
\end{equation}
we obtain the compliance tensor $s$. The stiffness tensor $c$ is
simply its inverse.  It should be noted that this method for the
calculation of elastic constants per atom can fail in the case of
microscopic plastic events.  We mostly avoid this by using a low
stress value, but for a small number of atoms the stiffness may be
unrealistically high or low.  The results we present later evidence
that this method is reliable enough for the present purpose.

While an amorphous system should have an isotropic shear modulus,
(partially) crystalline samples show a large anisotropy.  Earlier work
by Derlet and colleagues \cite{Derlet2012} demonstrated that the five
eigenshear moduli obtained from a Kelvin stiffness
tensor \cite{Thomson1856} are appropriate to characterize the elastic
moduli of amorphous systems. The Voigt notation of the stiffness tensor can
easily be converted to the Kelvin notation by the element-wise
multiplication
\begin{align}
  c_{ij}^\text{Kelvin} &= A_{ij} c_{ij}^\text{Voigt},
\end{align}
with
\begin{align}
  A &=
  \begin{pmatrix}
    1 & 1 & 1 & \sqrt{2} & \sqrt{2} & \sqrt{2} \\
    1 & 1 & 1 & \sqrt{2} & \sqrt{2} & \sqrt{2} \\
    1 & 1 & 1 & \sqrt{2} & \sqrt{2} & \sqrt{2} \\
    \sqrt{2} & \sqrt{2} & \sqrt{2} & 2 & 2 & 2 \\
    \sqrt{2} & \sqrt{2} & \sqrt{2} & 2 & 2 & 2 \\
    \sqrt{2} & \sqrt{2} & \sqrt{2} & 2 & 2 & 2 \\
  \end{pmatrix}.
\end{align}
We follow Derlet \textit{et al.} in calculating the shear moduli by
first projecting out the volume changes \cite{Derlet2012}:
\begin{align}
  c' &= P^\mathrm{T} \cdot c^\text{Kelvin} \cdot P
\end{align}
with
\begin{align}
  P &=
  \begin{pmatrix}
    +\frac{2}{3} & -\frac{1}{3} & -\frac{1}{3} & 0 & 0 & 0 \\[4pt]
    -\frac{1}{3} & +\frac{2}{3} & -\frac{1}{3} & 0 & 0 & 0 \\[4pt]
    -\frac{1}{3} & -\frac{1}{3} & +\frac{2}{3} & 0 & 0 & 0 \\[4pt]
    0 & 0 & 0 & 1 & 0 & 0 \\
    0 & 0 & 0 & 0 & 1 & 0 \\
    0 & 0 & 0 & 0 & 0 & 1
  \end{pmatrix}.
\end{align}
The shear moduli are now the five eigenvalues of $c'$, which---in an
isotropic material---correspond to two times the usual shear modulus
definition. A softening of the lowest shear modulus ($G_1$) was
connected to the boson peak modes \cite{Derlet2012}, which is why we
chose this value as our characteristic elastic constant.

\subsection{Structure analysis}

Structures and lattice defects were analyzed using \textsc{ovito}
\cite{Stukowski2010} and the algorithms implemented therein.
Crystalline ordering and its absence was identified using common
neighbor analysis (CNA) with an adaptive cutoff
\cite{Honeycutt1987,Stukowski2012}.  Atomic volumes were obtained by
Voronoi tessellation \cite{Voronoi1908, Voronoi1908a, Voronoi1909,
  Brostow1998}.

\section{Results}

\subsection{Defect-free high-entropy alloys}

\begin{figure}[b]
  \centering
  \includegraphics[]{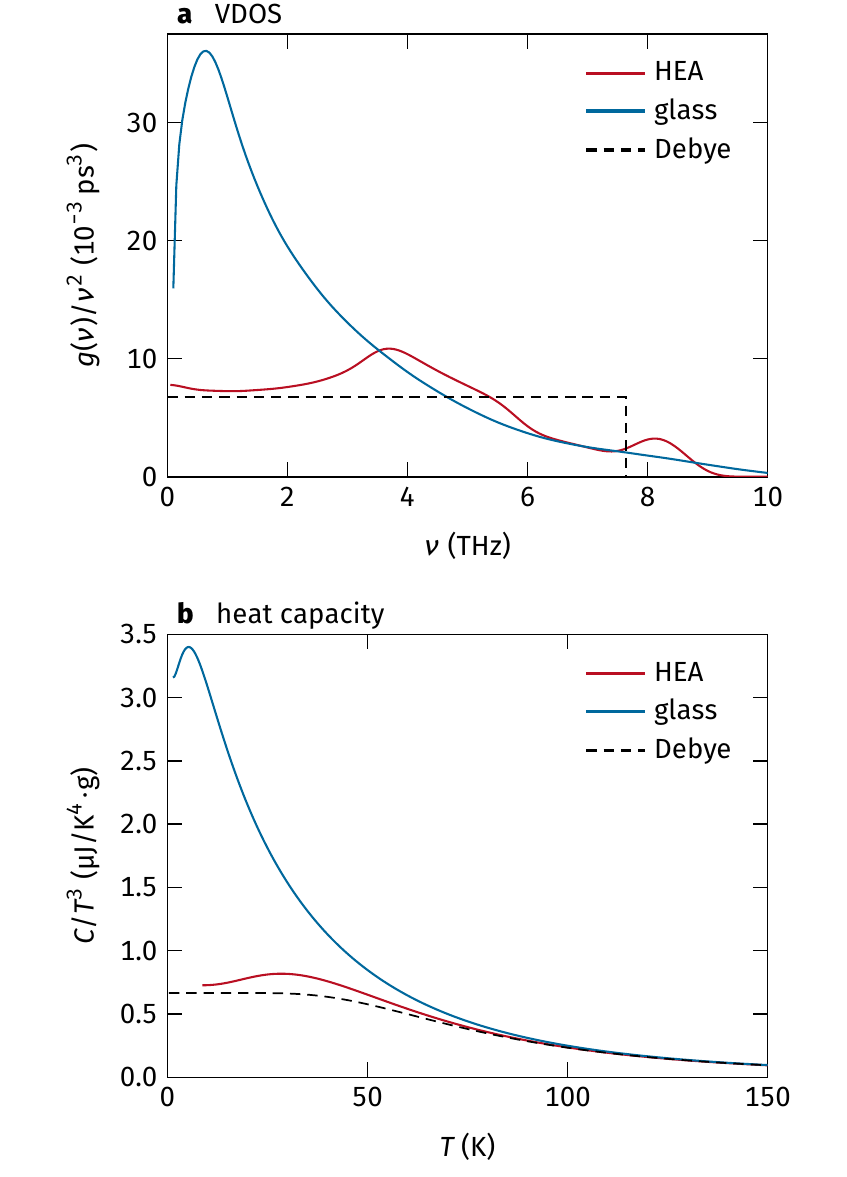}
  \caption{Vibrational spectrum and heat capacity of a high-entropy
    alloy compared with a glass.  (a) Reduced VDOS of the CuNiCoFe
    high-entropy alloy and a glass with the same composition. A clear
    boson peak can be detected only in the glass; the high-entropy
    alloy simply exhibits a van Hove singularity at \SI{4}{THz}. (b)
    The corresponding heat capacities as a function of temperature
    $T$. An excess between \SI{30}{K} and \SI{50}{K} is visible for
    the high-entropy alloy, which lies above the usual range for the
    boson peak of \SI{10}{K} to \SI{20}{K}.}
  \label{fig:VDOS-HEA-glass}
\end{figure}
In the first step, we consider a defect-free HEA at ambient pressure.
Figure~\ref{fig:VDOS-HEA-glass}(a) shows the VDOS $g$ as a function of
the frequency $\nu$, both for the HEA and the glass with the same
composition. A reduced depiction $g(\nu)/\nu^2$ over $\nu$ is chosen
in which the density of states postulated by the Debye model appears
as a constant and any excess over it is easily
visible. Figure~\ref{fig:VDOS-HEA-glass}(b) contains the corresponding
reduced heat capacities $C/T^3$. All curves feature an excess
contribution. However, in the HEA the vibrational excess is located at
around \SI{4}{THz}, while the boson peak in the glass appears at the
usual frequency of \SI{1}{THz}.

\begin{figure}
  \centering
  \includegraphics[]{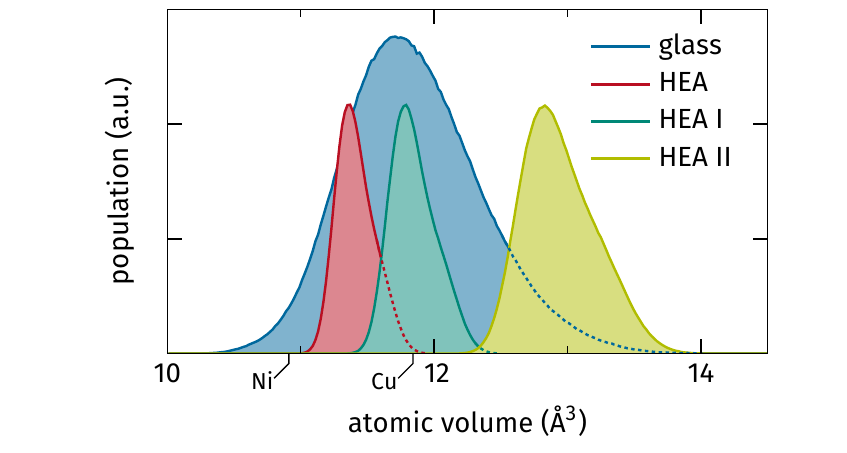}
  \caption{Atomic volume distribution at \SI{30}{K} of the
    high-entropy alloy compared with a glass of the same
    composition. The equilibrium HEA (red) is denser than the glass
    (blue) and lies between the densities of pure fcc nickel and
    copper---a small and a large constituent element of the alloy
    (indicated on the $x$ axis).  The sample HEA I is deformed
    elastically to obtain the same average density as the glass, while
    HEA II was deformed elastically to obtain densities similar to the
    largest glass atoms.}
  \label{fig:HEA-atvols}
\end{figure}
\begin{figure}
  \centering
  \includegraphics[]{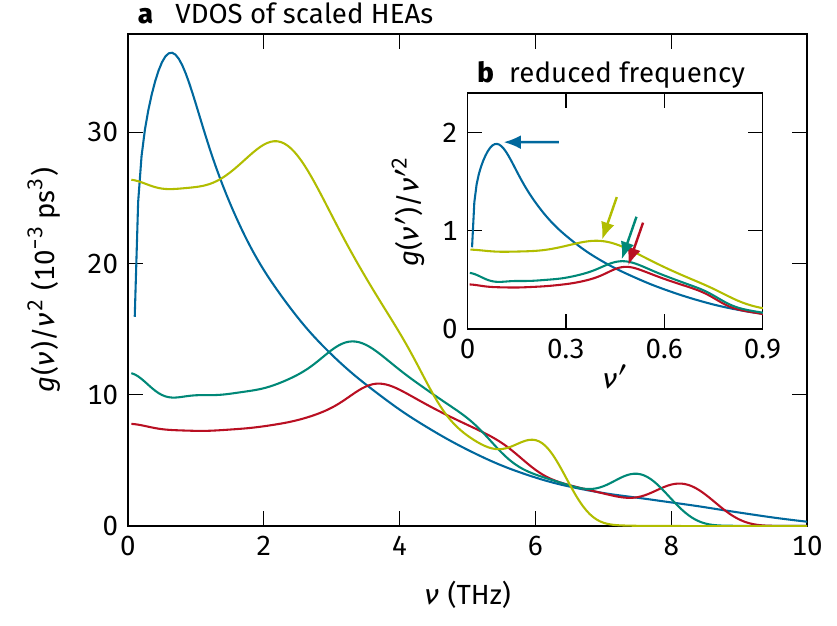}
  \caption{(a) Reduced VDOS of the glass compared with the three
    high-entropy alloy samples of different densities. The
    color coding matches that in Fig.~\ref{fig:HEA-atvols}. The
    inset (b) shows the same as a function of the reduced frequency
    $\nu' = \nu/\nu_\mathrm{D}$. These plots indicate that the van
    Hove singularity of the high-entropy alloys cannot be shifted by
    varying the density to obtain a boson peak.}
  \label{fig:VDOS-scaled-HEA}
\end{figure}
Given recent theories that the boson peak is a density-shifted van
Hove singularity and not disorder-related at all \cite{Chumakov2011},
it pays to have a more detailed look at the density dependence of the
VDOS. Figure~\ref{fig:HEA-atvols} contains an atomic volume histogram
of the HEA and the glass: Clearly, the HEA has a higher density. In
fact, the density of the HEA fluctuates between those of elemental Ni
and elemental Cu, a small and a large constituent element of the HEA.
Consequently, we first prepared a sample with a similar density to the
glass (HEA~I).  This was done by a simple, uniform scaling of the
lattice to reach the desired density followed by constant-volume
relaxation at \SI{30}{K}.  Recent computer simulations on Cu--Zr-based
metallic glasses demonstrate that only the atoms with the largest mean
square displacement---which make up 10\% of all atoms---contribute to
the boson peak \cite{Jakse2012}.  Assuming that those are the ones
with the highest atomic volume, we also prepared a sample which covers
atomic volumes among the highest in the glass (HEA~II).
The resulting VDOS measurements in Fig.~\ref{fig:VDOS-scaled-HEA}(a)
reveal that the glass VDOS is not reproduced even with a highly
strained HEA. While the boson peak is found (as usual) in the
\SI{1}{THz} region, the HEA peak only shifts to around
\SI{2}{THz}. Additionally, if plotting the reduced VDOS as a function
of a reduced frequency $\nu' = \nu/\nu_\mathrm{D}$, where
$\nu_\mathrm{D}$ is the Debye frequency, the HEA peaks fall more or
less on top of each other while the boson peak in the glass remains at
lower $\nu'$ [Fig.~\ref{fig:VDOS-scaled-HEA}(b)].

Thus, comparing the VDOS of various strained and unstrained HEAs with
a glass of the same composition does not provide evidence for a
relation of the boson peak to a shifted van Hove singularity in our
model alloy.  Moreover, the data suggests that the chemical disorder
of an HEA is insufficient to induce a boson peak.  This does not
exclude that lattices with stronger chemical disorder would show a
boson peak; indeed, several theoretical investigations use
fluctuations on a lattice to obtain these vibrational modes (see for
example Ref.~\onlinecite{Schirmacher1998}).  Such increased
disorder---which would most likely have to include elements with large
size differences---is expected to destabilize the simple fcc, bcc, or
hcp lattices of random metallic solid solutions \cite{Miracle2014, Inoue2000}
and is therefore not realizable in single-phase HEAs.

\subsection{Introduction of disorder via interstitial defects}

\begin{figure}
  \centering
  \includegraphics[]{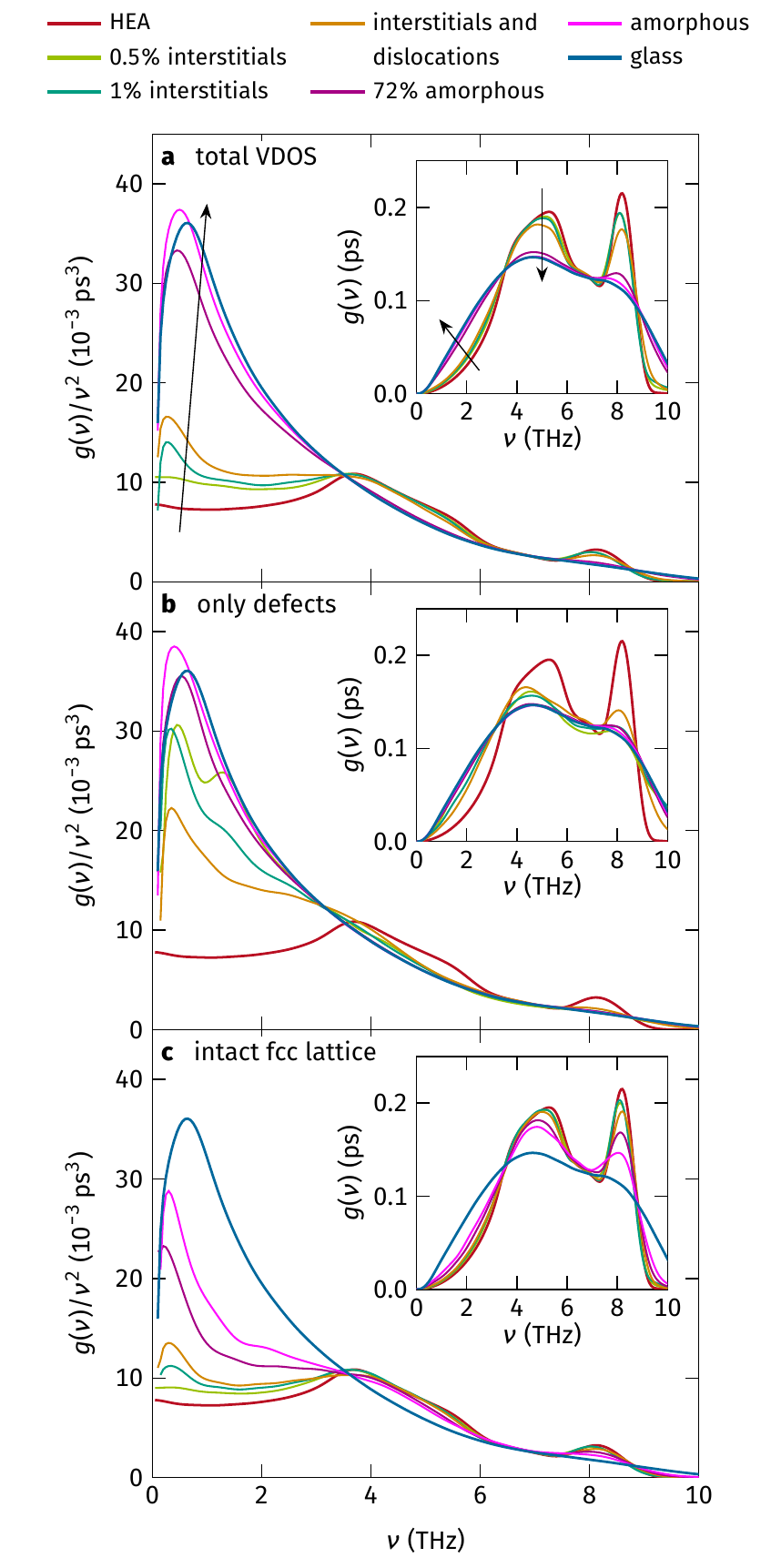}
  \caption{The effect of defects on the VDOS. (a) Total reduced VDOS
    for samples with different amounts of defects. The inset shows the
    nonreduced VDOS. The crystalline features stay visible but smear
    out with increasing defect concentration. The boson peak arises
    with the introduction of interstitials. (b) The partial VDOS of
    atoms with defective order, identified by CNA. Low concentrations
    of interstitials already lead to a significant boson peak
    signal. Still, the crystalline features persist for the defect
    atoms. (c) The partial VDOS of atoms on intact fcc lattice
    sites. The boson peak modes intrude at high defect concentrations,
    but the shape of the crystalline VDOS persists even for small fcc
    clusters.  The curves for the HEA and the glass always represent
    the total VDOS in every plot and are provided for
    comparison.}
  \label{fig:VDOS-defects}
\end{figure}

\begin{figure}
  \centering
  \includegraphics[]{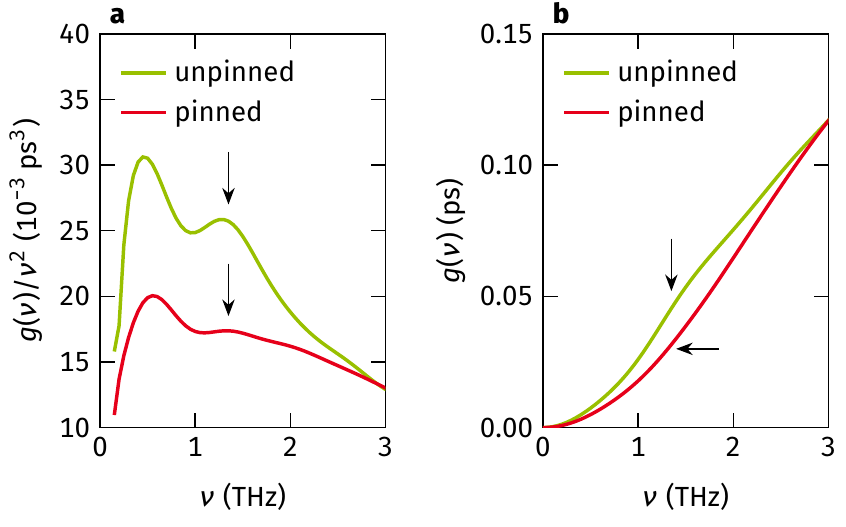}
  \caption{The origin of the split boson peak for the sample with
    0.5\% of interstitials: When pinning low-volume atoms ($\Omega <
    \SI{10.5}{\angstrom{}^3}$) which appear after the insertion of
    interstitials, the second peak is suppressed. This suggests that
    the splitting is a density-related frequency shift. The plots show
    the partial VDOS of only the defect atoms.}
  \label{fig:splitting}
\end{figure}
In the next step, we consider the influence of defects on the
vibrational spectrum since the boson peak was related to the lattice
dynamics of defects \cite{Schober2011} or loosely packed atoms
\cite{Schober1996, Li2006, Jakse2012} and subatomic voids
\cite{Sheng2012} in the past.  Interstitialcy theory states that
melting can be understood as the generation of an increasing number of
interstitials, and that, consequently, glasses are also characterized
by interstitial-like features \cite{Khonik2015}.  Following this, we
chose to randomly introduce interstitials into the HEA while keeping
the composition equimolar. After the insertion of various
concentrations of interstitials followed by equilibration at
\SI{30}{K} and ambient pressure, we obtained different structures:
samples with 0.5\% and 1\% interstitials, one sample in which most of
the point defects collapsed into stacking faults and dislocation-like
defects, one sample in which 72\% of atoms where amorphous, and one
sample which collapsed completely into an amorphous structure with
small crystalline clusters of a few atoms. The defects were identified
using CNA \cite{Honeycutt1987,
  Stukowski2012} in \textsc{ovito} \cite{Stukowski2010}.

\begin{figure*}
  \centering
  \includegraphics[]{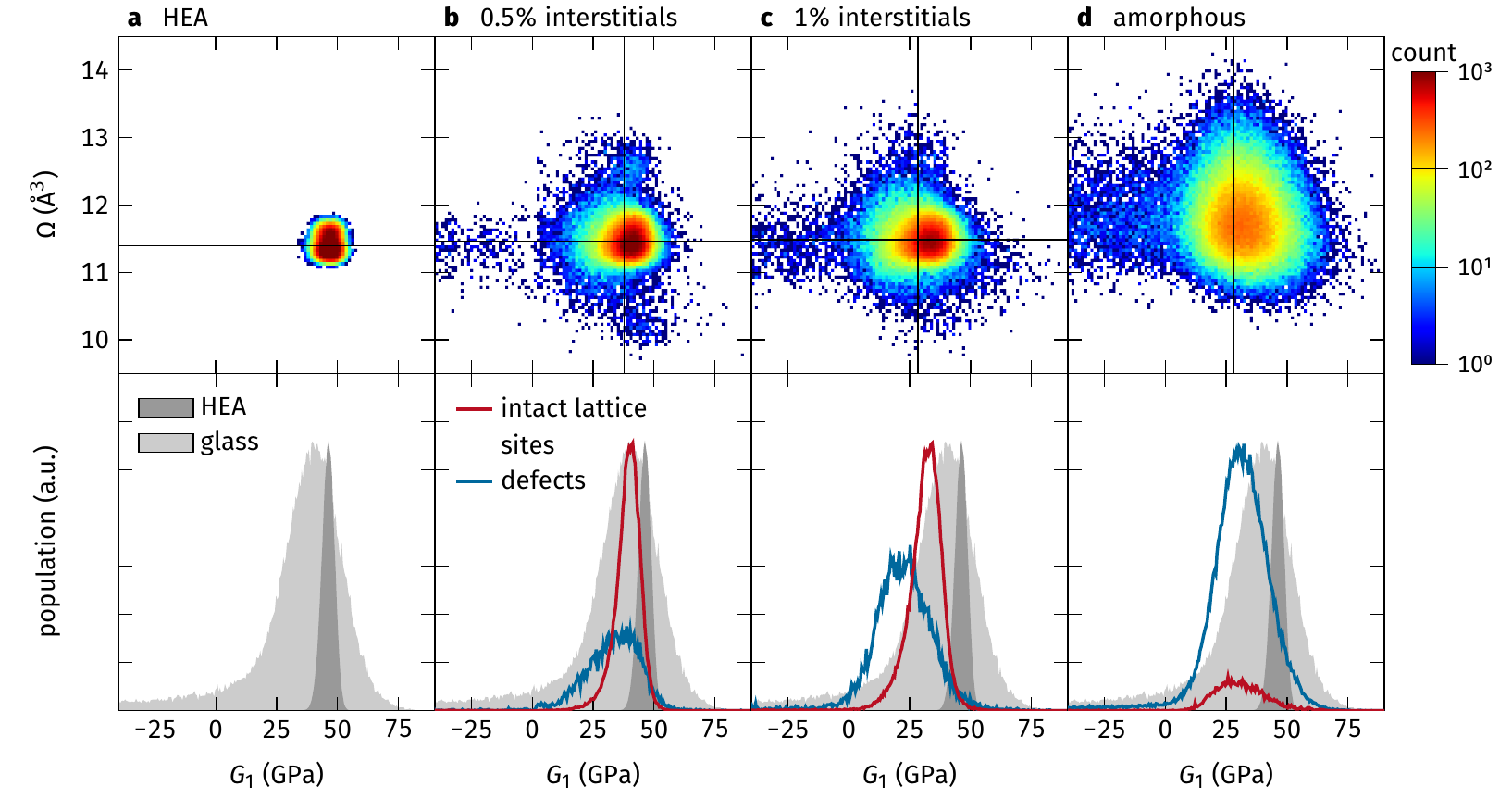}
  \caption{Histograms of atomic volume and per-atom shear moduli: (a)
    defect-free HEA; (b) HEA with 0.5\% interstitials; (c) HEA with
    1\% interstitials; (d) amorphous system.  The upper row shows 2D
    histograms of both the shear moduli $G_1$ and the atomic volumes
    $\Omega$ with a logarithmic color scale, while the lower row
    contains only the histograms of the shear moduli divided into
    defective atoms and atoms on intact fcc lattice sites.  The
    horizontal and vertical lines in the upper row represent the
    average values. The gray areas in the lower row show the
    distributions of the reference glass, as well as the defect-free
    HEA. The relative height of the distributions is not to scale in
    order to make small amounts of defective or lattice atoms visible.
    The HEA in (a) shows a narrow distribution on both axes. (b)--(d)
    With increasing defect concentration, regions with shear moduli
    below \SI{25}{GPa} emerge.  Those are due to the defect atoms,
    which have a shear modulus distribution similar to a bulk glass.
    (b) At low interstitial concentrations, atoms with increased and
    decreased volume appear. The decreased volume is obviously due to
    the insertion of interstitials, while the increased volume is
    unexpected. In fact, with increasing disorder in (c) and (d), the
    ``compressed'' atoms disappear. The reason for this is that the
    interstitials in the HEA always lead to local re-arrangements of
    the surrounding atoms, which in turn leads to a distribution of
    atomic volumes. The more the sample is defective, the easier the
    compressed atoms can release their stress and be accommodated in
    the disturbed lattice.}
  \label{fig:atvol-vs-G}
\end{figure*}
This increasing disorder now indeed leads to a boson peak as shown in
Fig.~\ref{fig:VDOS-defects}(a), where the peak height rises with the
defect concentration.  Contrary to other theoretical investigations
(for example Ref.~\onlinecite{Schirmacher2008}), the rise of this peak
is not correlated with a shift of the boson peak towards lower
frequencies. Instead, its position is more or less constant and even
shifts to somewhat higher frequencies when the lattice collapses.  We
split the VDOS into a partial vibrational density of states (PDOS) for
defective atoms surrounding point defects and atoms inside stacking
faults and amorphous regions [Fig.~\ref{fig:VDOS-defects}(b)], and a
PDOS for atoms on intact fcc lattice sites
[Fig.~\ref{fig:VDOS-defects}(c)].  The interstitials and surrounding
atoms already possess an almost full boson peak in the PDOS even at
small concentrations.  The contribution to the total VDOS is small
only because the intact fcc lattice sites, which are in the majority,
do not participate in the low-frequency modes.  The frequency shift is
even less pronounced in this plot.  The fact that the boson peak is
mostly absent in the PDOS of the intact lattice, especially at low
defect concentrations, demonstrates its localized nature.  This also
explains the frequency of the boson peak modes: The only difference
between the samples is the amount of defective areas. The disorder
inside these areas is comparable and therefore the boson peak modes
are located at the same frequencies.  The inset of
Fig.~\ref{fig:VDOS-defects}(a) explains the additional small frequency
shift of the samples which were at least partially amorphous: The
features of the crystalline VDOS smear out upon amorphization and the
low-frequency tail of the first crystalline peak overlaps with the
boson peak, leading to an apparent shift to higher frequencies.

At 0.5\% of interstitials, a second peak slightly below \SI{2}{THz}
appears.  It stands to reason that the introduction of interstitials
leads to a compression of some neighboring atoms. While a chemically
ordered system would only exhibit compressed atoms, the HEA locally
rearranges and produces both atoms with increased and reduced atomic
volume. Such a density difference could lead to a frequency shift
between both types of defective atoms, which would in turn lead to a
splitting of the boson peak. We tested this by pinning high-density
atoms and recomputing the VDOS of the non-pinned defect atoms.
Figure~\ref{fig:splitting} shows that this indeed switches off the
\SI{2}{THz} peak.  At higher defect concentrations the split
disappears. As we will discuss later in some more detail, the
compressed atoms disappear with increasing defect concentration, as
they are accommodated in the defective parts of the lattice.

The defect PDOS of the sample which contains both interstitials and
dislocations seems to be reduced again, which indicates that the
dislocation and stacking-fault defects do not contribute to the boson
peak, thereby reducing the combined value.

Finally, the collapsed lattice has a VDOS similar to the glass, except
for a small shift to lower frequencies. This can be explained by the
fact that the collapsed glass did not have time for structural
relaxation. This relaxation would reduce the density and consequently
lead to a slight shift to higher frequencies. In the samples with such
a high number of defects, the PDOS of the intact lattice sites also
exhibit low-frequency modes. This is because a large fraction of these
atoms has disordered neighbors, the vibrations of which also affect
the fcc-ordered atoms. Still, even small fcc clusters resist a
complete intrusion of the boson peak.

\begin{figure*}
  \centering
  \includegraphics[]{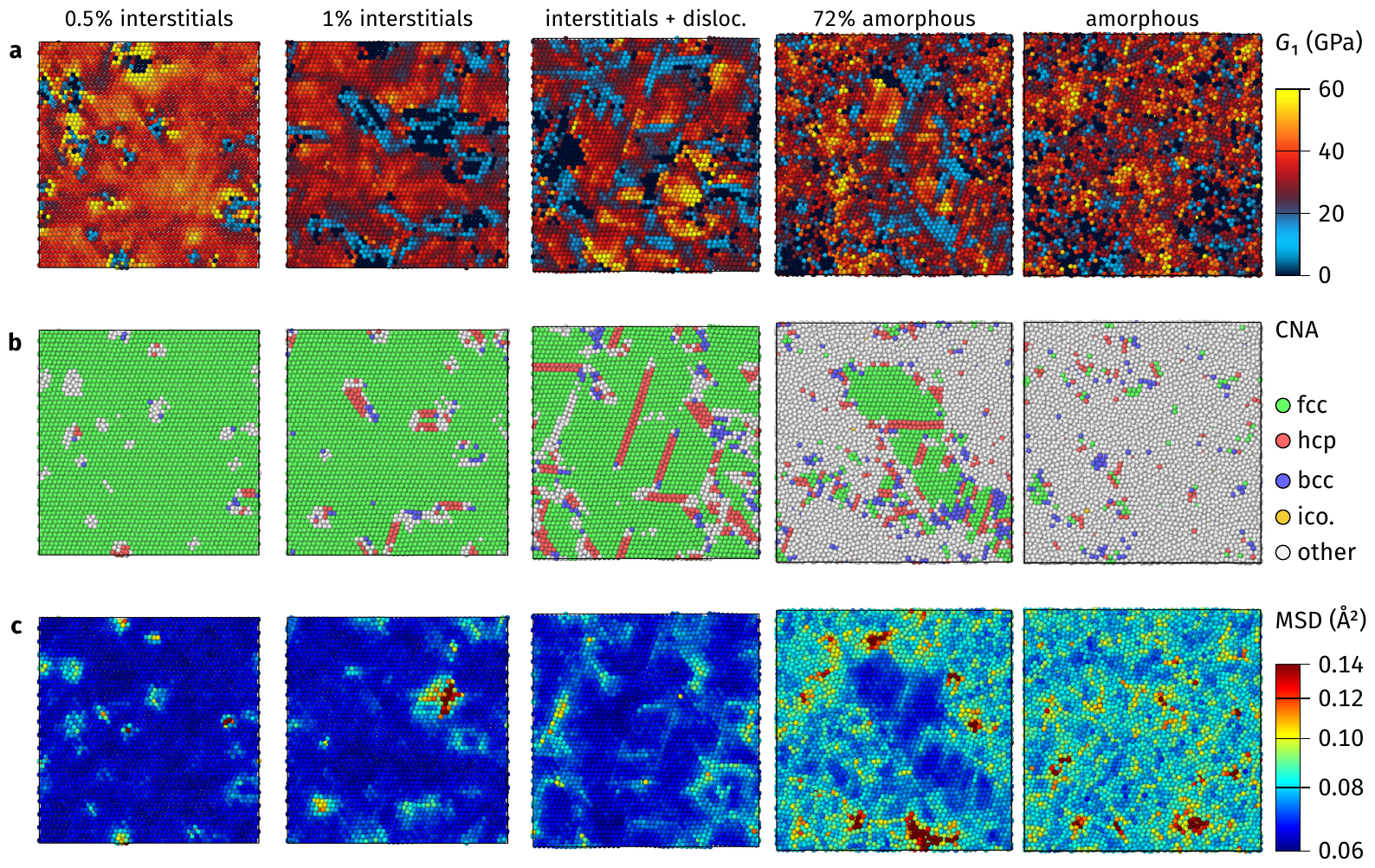}
  \caption{Snapshots of samples with different defects, mapping the
    spatial distribution of shear modulus, disorder, and mean square
    displacement. A correlation between low shear modulus (a), defects
    as identified by CNA (b), and high mean square displacement (c) is
    apparent. The color scale in (c) is logarithmic to enhance the
    visibility of small displacements.}
  \label{fig:maps}
\end{figure*}
These results support the view that the boson peak is due to localized
defect-related modes. A shift of a van Hove singularity can now be
completely excluded as even the defect atoms still retain the VDOS
shape of the crystal, although heavily smeared out. The boson peak
arises as additional modes around the interstitials.

\subsection{Connection between defective structures and softening}
While we excluded that the fluctuations of elastic constants of the
HEA are sufficient to induce a boson peak, it pays to have another
look at the shear moduli of the defective samples.  Following Derlet
\textit{et al.}, we use the lowest of the five Kelvin shear moduli,
$G_1$, which has been related to the boson peak \cite{Derlet2012}.
Two-dimensional (2D) histograms of the atomic volume $\Omega$ versus
the per-atom shear modulus $G_1$ are presented in
Fig.~\ref{fig:atvol-vs-G}. Although the HEA has fluctuations in both,
they are small compared to the defective samples. Already at low
interstitial concentrations, significant amounts of atoms with moduli
between \SI{0}{GPa} and \SI{25}{GPa} appear.  With increasing defect
density the atoms with low modulus become more numerous.  Thus,
significant softening is needed for the boson peak, if it in fact
results from fluctuating moduli.  One can observe a general softening
of the whole material with rising interstitial concentration, which
points to a generally increasing instability of the material. This is
to be expected at such high defect concentrations. The amorphous
system is slightly softer than the quenched glass, which supports our
earlier argument that the quenched glass represents a more relaxed
state. Qualitatively, the two systems are similar.
Figure~\ref{fig:atvol-vs-G}(b) also proves our earlier statement that
atoms with increased as well as reduced density occur.  As we can see
in Figs.~\ref{fig:atvol-vs-G}(c) and \ref{fig:atvol-vs-G}(d), the
unfavorable high-density atoms start disappearing because the systems
with higher defect concentrations can re-arrange more easily. With
them, the split in the boson peak also disappears.

The remaining question is, if the defect-based boson peak models are
indeed as different as they initially appear to be. Studies
ascribe the boson peak to interstitialcies \cite{Granato1996,
  Vasiliev2009}, ``liquidlike'' regions \cite{Sheng2012, Ding2014b},
``rattling'' atoms \cite{Guerdane2008, Jakse2012}, and of course
fluctuating force constants \cite{Schirmacher1993, Schirmacher1998,
  Schirmacher2007, Marruzzo2013, Schirmacher2015}.  So, in addition to
local shear moduli, we calculated the mean square displacement (MSD)
of the atoms at \SI{30}{K} over a period of \SI{100}{ps}.  Shear
modulus $G_1$, structure as identified by CNA, and MSD are shown in
Fig.~\ref{fig:maps}. The spatial correlation between defect atoms, low
shear modulus, and high MSD is apparent. This is not surprising: The
defects introduced into the lattice are well known to soften the
material \cite{Khonik2015}.  Furthermore, low moduli are of course
connected with shallow potential wells and therefore a high MSD. This
suggests that softened regions can be introduced by interstitials, and
that these regions resemble the ``soft spots'' discussed in the
literature on metallic glasses \cite{Sheng2012, Ding2014b}.  Thus,
previous studies seem to observe the same effect by measuring
different quantities.

\subsection{A boson peak without chemical disorder}
\begin{figure}
  \centering
  \includegraphics[]{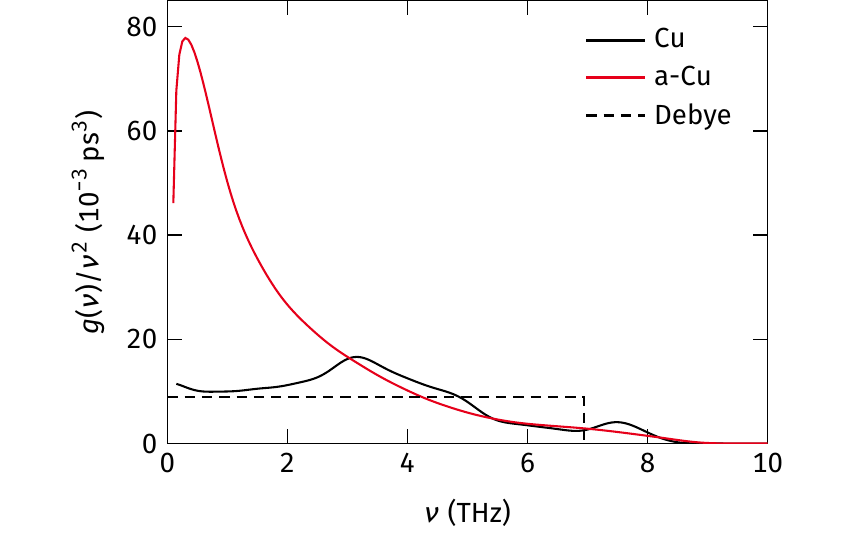}
  \caption{Comparison of the VDOS of fcc copper with amorphous
    copper. A boson peak occurs even without chemical disorder. The
    VDOS of crystalline copper resembles that of the HEA and is
    typical for fcc metals. The peak around \SI{3.5}{THz} is a van
    Hove singularity.}
  \label{fig:VDOS-Cu}
\end{figure}
This analysis again supports the view that the loss of structural
order is the key ingredient for obtaining the boson peak. In order to
supplement this picture, we went one step further and removed the
chemical disorder from the model structure by studying a
single-component glass. We repeated the analysis presented in
Fig.~\ref{fig:VDOS-HEA-glass} for fcc and amorphous copper (a-Cu)
quenched from the melt with \SI{e14}{K/s}. As can be seen in
Fig.~\ref{fig:VDOS-Cu}, the a-Cu sample also exhibits an excess of
states at low frequencies. This evidences that chemical disorder is
not only not a sufficient condition for the boson peak, but that it is
not even a necessary one. Indeed, the shape of the VDOS of fcc copper
resembles that of the HEA. The excess states in the HEA are simply
due to a van Hove singularity. Contrary to recent theories
\cite{Chumakov2011, Chumakov2014, Chumakov2015, Chumakov2016,
  Baldi2016}, this van Hove singularity is not the same as a boson
peak. As demonstrated above, (i) the van Hove singularity does not
shift to \SI{1}{THz} when scaling the density to values typical for
the glass, and (ii) the crystalline features of the VDOS also appear
in the glass; heavily smeared out but unshifted.

\section{Conclusion}

All in all, the current work supports the conclusion that the boson
peak in alloys results from \mbox{(quasi-)}localized additional modes.
The vibrational modes of interstitial defects resemble those of a
glass and the atomic volumes around those defects do not exceed the
values in HEA~II, a defect-free lattice with reduced density.  This
excludes a simple density-related origin of these additional modes in
the system we considered, since HEA~II did not exhibit a boson peak.
In fact, the resemblance of the PDOS in Fig.~\ref{fig:VDOS-defects}(b)
to the crystalline state excludes a general shift of frequencies.
This is also in accordance with recent measurements and simulations of
the boson peak in deformed Cu--Zr and Pd--Ni--P glasses \cite{Bunz2014,
  Mitrofanov2014}.  The boson peak contribution inside the shear band
was found to be higher than in the undeformed material. While one
could argue that the density in the shear band is
lowered \cite{Ritter2011, Ritter2012}, the boson peak does not shift
very much on the frequency or temperature axis. Together with the
current results, and the fact that we know that short-range order
inside the shear bands is disturbed \cite{Ritter2011, Ritter2012},
this data rather supports the defect picture outlined above.
The results presented here also confirm what was suggested by
Schirmacher and colleagues \cite{Schirmacher2015}: Defect-based
pictures of the boson peak origin can be united with the fluctuating
force-constant model, although our results indicate that a significant
localized softening is needed.  The ``soft spot'' picture of metallic
glasses seems relevant here: Regions of defective short-range order
are connected with lower stiffness and strength as well as with a high
boson peak signal \cite{Tanguy2010, Sheng2012, Derlet2012, Ding2014b}.
Of course, atoms in these regions exhibit high MSDs as was shown in
previous studies \cite{Guerdane2008, Jakse2012}. This suggests a
relevance of the boson peak not only in the realm of solid-state
physics, but also for the mechanical properties of metallic glasses.

\begin{acknowledgments}
The authors thank Alexander Stukowski for help with \textsc{ovito},
Peter Derlet for pointing us to the Kelvin stiffness tensor, as well
as Gerhard Wilde and Sergiy Divinski for discussions about the
low-temperature heat capacity of HEAs.
Financial support by the Deutsche For\-schungs\-ge\-mein\-schaft (DFG)
through Grant No.\ AL \mbox{578/6-2}, as well as a travel grant to
Finland through the PPP program of the Deut\-scher Aka\-de\-mi\-scher
Aus\-tausch\-dienst (DAAD) are gratefully acknowledged.  Computing
time was made available by the Technische Universit\"at Darmstadt on
the Lichtenberg cluster.
\end{acknowledgments}


\begin{thebibliography}{58}%
\makeatletter
\providecommand \@ifxundefined [1]{%
 \@ifx{#1\undefined}
}%
\providecommand \@ifnum [1]{%
 \ifnum #1\expandafter \@firstoftwo
 \else \expandafter \@secondoftwo
 \fi
}%
\providecommand \@ifx [1]{%
 \ifx #1\expandafter \@firstoftwo
 \else \expandafter \@secondoftwo
 \fi
}%
\providecommand \natexlab [1]{#1}%
\providecommand \enquote  [1]{``#1''}%
\providecommand \bibnamefont  [1]{#1}%
\providecommand \bibfnamefont [1]{#1}%
\providecommand \citenamefont [1]{#1}%
\providecommand \href@noop [0]{\@secondoftwo}%
\providecommand \href [0]{\begingroup \@sanitize@url \@href}%
\providecommand \@href[1]{\@@startlink{#1}\@@href}%
\providecommand \@@href[1]{\endgroup#1\@@endlink}%
\providecommand \@sanitize@url [0]{\catcode `\\12\catcode `\$12\catcode
  `\&12\catcode `\#12\catcode `\^12\catcode `\_12\catcode `\%12\relax}%
\providecommand \@@startlink[1]{}%
\providecommand \@@endlink[0]{}%
\providecommand \url  [0]{\begingroup\@sanitize@url \@url }%
\providecommand \@url [1]{\endgroup\@href {#1}{\urlprefix }}%
\providecommand \urlprefix  [0]{URL }%
\providecommand \Eprint [0]{\href }%
\providecommand \doibase [0]{http://dx.doi.org/}%
\providecommand \selectlanguage [0]{\@gobble}%
\providecommand \bibinfo  [0]{\@secondoftwo}%
\providecommand \bibfield  [0]{\@secondoftwo}%
\providecommand \translation [1]{[#1]}%
\providecommand \BibitemOpen [0]{}%
\providecommand \bibitemStop [0]{}%
\providecommand \bibitemNoStop [0]{.\EOS\space}%
\providecommand \EOS [0]{\spacefactor3000\relax}%
\providecommand \BibitemShut  [1]{\csname bibitem#1\endcsname}%
\let\auto@bib@innerbib\@empty
\bibitem [{\citenamefont {Elliott}(1992)}]{Elliott1992}%
  \BibitemOpen
  \bibfield  {author} {\bibinfo {author} {\bibfnamefont {S.~R.}\ \bibnamefont
  {Elliott}},\ }\href {http://dx.doi.org/10.1209/0295-5075/19/3/009} {\bibfield
   {journal} {\bibinfo  {journal} {EPL}\ }\textbf {\bibinfo {volume} {19}},\
  \bibinfo {pages} {201} (\bibinfo {year} {1992})}\BibitemShut {NoStop}%
\bibitem [{\citenamefont {Karpov}\ \emph {et~al.}(1983)\citenamefont {Karpov},
  \citenamefont {Klinger},\ and\ \citenamefont {Ignat'ev}}]{Karpov1983}%
  \BibitemOpen
  \bibfield  {author} {\bibinfo {author} {\bibfnamefont {V.~G.}\ \bibnamefont
  {Karpov}}, \bibinfo {author} {\bibfnamefont {M.~I.}\ \bibnamefont {Klinger}},
  \ and\ \bibinfo {author} {\bibfnamefont {F.~N.}\ \bibnamefont {Ignat'ev}},\
  }\href {http://jetp.ac.ru/cgi-bin/e/index/e/57/2/p439?a=list} {\bibfield
  {journal} {\bibinfo  {journal} {Zh. Eksp. Teor. Fiz.}\ }\textbf {\bibinfo
  {volume} {84}},\ \bibinfo {pages} {760} (\bibinfo {year} {1983})}\BibitemShut
  {NoStop}%
\bibitem [{\citenamefont {Laird}\ and\ \citenamefont
  {Schober}(1991)}]{Laird1991}%
  \BibitemOpen
  \bibfield  {author} {\bibinfo {author} {\bibfnamefont {B.~B.}\ \bibnamefont
  {Laird}}\ and\ \bibinfo {author} {\bibfnamefont {H.~R.}\ \bibnamefont
  {Schober}},\ }\href {\doibase 10.1103/PhysRevLett.66.636} {\bibfield
  {journal} {\bibinfo  {journal} {Phys. Rev. Lett.}\ }\textbf {\bibinfo
  {volume} {66}},\ \bibinfo {pages} {636} (\bibinfo {year} {1991})}\BibitemShut
  {NoStop}%
\bibitem [{\citenamefont {Schober}\ and\ \citenamefont
  {Laird}(1991)}]{Schober1991}%
  \BibitemOpen
  \bibfield  {author} {\bibinfo {author} {\bibfnamefont {H.~R.}\ \bibnamefont
  {Schober}}\ and\ \bibinfo {author} {\bibfnamefont {B.~B.}\ \bibnamefont
  {Laird}},\ }\href {\doibase 10.1103/PhysRevB.44.6746} {\bibfield  {journal}
  {\bibinfo  {journal} {Phys. Rev. B}\ }\textbf {\bibinfo {volume} {44}},\
  \bibinfo {pages} {6746} (\bibinfo {year} {1991})}\BibitemShut {NoStop}%
\bibitem [{\citenamefont {Gil}\ \emph {et~al.}(1993)\citenamefont {Gil},
  \citenamefont {Ramos}, \citenamefont {Bringer},\ and\ \citenamefont
  {Buchenau}}]{Gil1993}%
  \BibitemOpen
  \bibfield  {author} {\bibinfo {author} {\bibfnamefont {L.}~\bibnamefont
  {Gil}}, \bibinfo {author} {\bibfnamefont {M.~A.}\ \bibnamefont {Ramos}},
  \bibinfo {author} {\bibfnamefont {A.}~\bibnamefont {Bringer}}, \ and\
  \bibinfo {author} {\bibfnamefont {U.}~\bibnamefont {Buchenau}},\ }\href
  {\doibase 10.1103/PhysRevLett.70.182} {\bibfield  {journal} {\bibinfo
  {journal} {Phys. Rev. Lett.}\ }\textbf {\bibinfo {volume} {70}},\ \bibinfo
  {pages} {182} (\bibinfo {year} {1993})}\BibitemShut {NoStop}%
\bibitem [{\citenamefont {Schober}\ and\ \citenamefont
  {Oligschleger}(1996)}]{Schober1996}%
  \BibitemOpen
  \bibfield  {author} {\bibinfo {author} {\bibfnamefont {H.~R.}\ \bibnamefont
  {Schober}}\ and\ \bibinfo {author} {\bibfnamefont {C.}~\bibnamefont
  {Oligschleger}},\ }\href {\doibase 10.1103/PhysRevB.53.11469} {\bibfield
  {journal} {\bibinfo  {journal} {Phys. Rev. B}\ }\textbf {\bibinfo {volume}
  {53}},\ \bibinfo {pages} {11469} (\bibinfo {year} {1996})}\BibitemShut
  {NoStop}%
\bibitem [{\citenamefont {Shintani}\ and\ \citenamefont
  {Tanaka}(2008)}]{Shintani2008}%
  \BibitemOpen
  \bibfield  {author} {\bibinfo {author} {\bibfnamefont {H.}~\bibnamefont
  {Shintani}}\ and\ \bibinfo {author} {\bibfnamefont {H.}~\bibnamefont
  {Tanaka}},\ }\href {\doibase 10.1038/nmat2293} {\bibfield  {journal}
  {\bibinfo  {journal} {Nat. Mater.}\ }\textbf {\bibinfo {volume} {7}},\
  \bibinfo {pages} {870} (\bibinfo {year} {2008})}\BibitemShut {NoStop}%
\bibitem [{\citenamefont {Schober}(2011)}]{Schober2011}%
  \BibitemOpen
  \bibfield  {author} {\bibinfo {author} {\bibfnamefont {H.~R.}\ \bibnamefont
  {Schober}},\ }\href {\doibase 10.1016/j.jnoncrysol.2010.07.036} {\bibfield
  {journal} {\bibinfo  {journal} {J. Non-Cryst. Solids}\ }\textbf {\bibinfo
  {volume} {357}},\ \bibinfo {pages} {501} (\bibinfo {year}
  {2011})}\BibitemShut {NoStop}%
\bibitem [{\citenamefont {Schober}\ \emph {et~al.}(2014)\citenamefont
  {Schober}, \citenamefont {Buchenau},\ and\ \citenamefont
  {Gurevich}}]{Schober2014}%
  \BibitemOpen
  \bibfield  {author} {\bibinfo {author} {\bibfnamefont {H.~R.}\ \bibnamefont
  {Schober}}, \bibinfo {author} {\bibfnamefont {U.}~\bibnamefont {Buchenau}}, \
  and\ \bibinfo {author} {\bibfnamefont {V.~L.}\ \bibnamefont {Gurevich}},\
  }\href {\doibase 10.1103/PhysRevB.89.014204} {\bibfield  {journal} {\bibinfo
  {journal} {Phys. Rev. B}\ }\textbf {\bibinfo {volume} {89}},\ \bibinfo
  {pages} {014204} (\bibinfo {year} {2014})}\BibitemShut {NoStop}%
\bibitem [{\citenamefont {Li}\ \emph {et~al.}(2006)\citenamefont {Li},
  \citenamefont {Bai}, \citenamefont {Wang},\ and\ \citenamefont
  {Samwer}}]{Li2006}%
  \BibitemOpen
  \bibfield  {author} {\bibinfo {author} {\bibfnamefont {Y.}~\bibnamefont
  {Li}}, \bibinfo {author} {\bibfnamefont {H.~Y.}\ \bibnamefont {Bai}},
  \bibinfo {author} {\bibfnamefont {W.~H.}\ \bibnamefont {Wang}}, \ and\
  \bibinfo {author} {\bibfnamefont {K.}~\bibnamefont {Samwer}},\ }\href
  {\doibase 10.1103/PhysRevB.74.052201} {\bibfield  {journal} {\bibinfo
  {journal} {Phys. Rev. B}\ }\textbf {\bibinfo {volume} {74}},\ \bibinfo
  {pages} {052201} (\bibinfo {year} {2006})}\BibitemShut {NoStop}%
\bibitem [{\citenamefont {Guerdane}\ and\ \citenamefont
  {Teichler}(2008)}]{Guerdane2008}%
  \BibitemOpen
  \bibfield  {author} {\bibinfo {author} {\bibfnamefont {M.}~\bibnamefont
  {Guerdane}}\ and\ \bibinfo {author} {\bibfnamefont {H.}~\bibnamefont
  {Teichler}},\ }\href {\doibase 10.1103/PhysRevLett.101.065506} {\bibfield
  {journal} {\bibinfo  {journal} {Phys. Rev. Lett.}\ }\textbf {\bibinfo
  {volume} {101}},\ \bibinfo {pages} {065506} (\bibinfo {year}
  {2008})}\BibitemShut {NoStop}%
\bibitem [{\citenamefont {Jakse}\ \emph {et~al.}(2012)\citenamefont {Jakse},
  \citenamefont {Nassour},\ and\ \citenamefont {Pasturel}}]{Jakse2012}%
  \BibitemOpen
  \bibfield  {author} {\bibinfo {author} {\bibfnamefont {N.}~\bibnamefont
  {Jakse}}, \bibinfo {author} {\bibfnamefont {A.}~\bibnamefont {Nassour}}, \
  and\ \bibinfo {author} {\bibfnamefont {A.}~\bibnamefont {Pasturel}},\ }\href
  {\doibase 10.1103/PhysRevB.85.174201} {\bibfield  {journal} {\bibinfo
  {journal} {Phys. Rev. B}\ }\textbf {\bibinfo {volume} {85}},\ \bibinfo
  {pages} {174201} (\bibinfo {year} {2012})}\BibitemShut {NoStop}%
\bibitem [{\citenamefont {Sheng}\ \emph {et~al.}(2012)\citenamefont {Sheng},
  \citenamefont {Ma},\ and\ \citenamefont {Kramer}}]{Sheng2012}%
  \BibitemOpen
  \bibfield  {author} {\bibinfo {author} {\bibfnamefont {H.}~\bibnamefont
  {Sheng}}, \bibinfo {author} {\bibfnamefont {E.}~\bibnamefont {Ma}}, \ and\
  \bibinfo {author} {\bibfnamefont {M.}~\bibnamefont {Kramer}},\ }\href
  {\doibase 10.1007/s11837-012-0360-y} {\bibfield  {journal} {\bibinfo
  {journal} {JOM}\ }\textbf {\bibinfo {volume} {64}},\ \bibinfo {pages} {856}
  (\bibinfo {year} {2012})}\BibitemShut {NoStop}%
\bibitem [{\citenamefont {Granato}(1996)}]{Granato1996}%
  \BibitemOpen
  \bibfield  {author} {\bibinfo {author} {\bibfnamefont {A.~V.}\ \bibnamefont
  {Granato}},\ }\href {\doibase 10.1016/0921-4526(95)00716-4} {\bibfield
  {journal} {\bibinfo  {journal} {Physica B: Condens. Matter}\ }\textbf
  {\bibinfo {volume} {219--220}},\ \bibinfo {pages} {270} (\bibinfo {year}
  {1996})}\BibitemShut {NoStop}%
\bibitem [{\citenamefont {Vasiliev}\ \emph {et~al.}(2009)\citenamefont
  {Vasiliev}, \citenamefont {Voloshok}, \citenamefont {Granato}, \citenamefont
  {Joncich}, \citenamefont {Mitrofanov},\ and\ \citenamefont
  {Khonik}}]{Vasiliev2009}%
  \BibitemOpen
  \bibfield  {author} {\bibinfo {author} {\bibfnamefont {A.~N.}\ \bibnamefont
  {Vasiliev}}, \bibinfo {author} {\bibfnamefont {T.~N.}\ \bibnamefont
  {Voloshok}}, \bibinfo {author} {\bibfnamefont {A.~V.}\ \bibnamefont
  {Granato}}, \bibinfo {author} {\bibfnamefont {D.~M.}\ \bibnamefont
  {Joncich}}, \bibinfo {author} {\bibfnamefont {Y.~P.}\ \bibnamefont
  {Mitrofanov}}, \ and\ \bibinfo {author} {\bibfnamefont {V.~A.}\ \bibnamefont
  {Khonik}},\ }\href {\doibase 10.1103/PhysRevB.80.172102} {\bibfield
  {journal} {\bibinfo  {journal} {Phys. Rev. B}\ }\textbf {\bibinfo {volume}
  {80}},\ \bibinfo {pages} {172102} (\bibinfo {year} {2009})}\BibitemShut
  {NoStop}%
\bibitem [{\citenamefont {Grigera}\ \emph {et~al.}(2003)\citenamefont
  {Grigera}, \citenamefont {Mart\'in-Mayor}, \citenamefont {Parisi},\ and\
  \citenamefont {Verrocchio}}]{Grigera2003}%
  \BibitemOpen
  \bibfield  {author} {\bibinfo {author} {\bibfnamefont {T.~S.}\ \bibnamefont
  {Grigera}}, \bibinfo {author} {\bibfnamefont {V.}~\bibnamefont
  {Mart\'in-Mayor}}, \bibinfo {author} {\bibfnamefont {G.}~\bibnamefont
  {Parisi}}, \ and\ \bibinfo {author} {\bibfnamefont {P.}~\bibnamefont
  {Verrocchio}},\ }\href {\doibase 10.1038/nature01475} {\bibfield  {journal}
  {\bibinfo  {journal} {Nature}\ }\textbf {\bibinfo {volume} {422}},\ \bibinfo
  {pages} {289} (\bibinfo {year} {2003})}\BibitemShut {NoStop}%
\bibitem [{\citenamefont {Schirmacher}\ and\ \citenamefont
  {Wagener}(1993)}]{Schirmacher1993}%
  \BibitemOpen
  \bibfield  {author} {\bibinfo {author} {\bibfnamefont {W.}~\bibnamefont
  {Schirmacher}}\ and\ \bibinfo {author} {\bibfnamefont {M.}~\bibnamefont
  {Wagener}},\ }\href {\doibase 10.1016/0038-1098(93)90147-F} {\bibfield
  {journal} {\bibinfo  {journal} {Solid State Commun.}\ }\textbf {\bibinfo
  {volume} {86}},\ \bibinfo {pages} {597} (\bibinfo {year} {1993})}\BibitemShut
  {NoStop}%
\bibitem [{\citenamefont {Schirmacher}\ \emph {et~al.}(1998)\citenamefont
  {Schirmacher}, \citenamefont {Diezemann},\ and\ \citenamefont
  {Ganter}}]{Schirmacher1998}%
  \BibitemOpen
  \bibfield  {author} {\bibinfo {author} {\bibfnamefont {W.}~\bibnamefont
  {Schirmacher}}, \bibinfo {author} {\bibfnamefont {G.}~\bibnamefont
  {Diezemann}}, \ and\ \bibinfo {author} {\bibfnamefont {C.}~\bibnamefont
  {Ganter}},\ }\href {\doibase 10.1103/PhysRevLett.81.136} {\bibfield
  {journal} {\bibinfo  {journal} {Phys. Rev. Lett.}\ }\textbf {\bibinfo
  {volume} {81}},\ \bibinfo {pages} {136} (\bibinfo {year} {1998})}\BibitemShut
  {NoStop}%
\bibitem [{\citenamefont {Schirmacher}\ \emph {et~al.}(2007)\citenamefont
  {Schirmacher}, \citenamefont {Ruocco},\ and\ \citenamefont
  {Scopigno}}]{Schirmacher2007}%
  \BibitemOpen
  \bibfield  {author} {\bibinfo {author} {\bibfnamefont {W.}~\bibnamefont
  {Schirmacher}}, \bibinfo {author} {\bibfnamefont {G.}~\bibnamefont {Ruocco}},
  \ and\ \bibinfo {author} {\bibfnamefont {T.}~\bibnamefont {Scopigno}},\
  }\href {\doibase 10.1103/PhysRevLett.98.025501} {\bibfield  {journal}
  {\bibinfo  {journal} {Phys. Rev. Lett.}\ }\textbf {\bibinfo {volume} {98}},\
  \bibinfo {pages} {025501} (\bibinfo {year} {2007})}\BibitemShut {NoStop}%
\bibitem [{\citenamefont {Schirmacher}\ \emph {et~al.}(2008)\citenamefont
  {Schirmacher}, \citenamefont {Schmid}, \citenamefont {Tomaras}, \citenamefont
  {Viliani}, \citenamefont {Baldi}, \citenamefont {Ruocco},\ and\ \citenamefont
  {Scopigno}}]{Schirmacher2008}%
  \BibitemOpen
  \bibfield  {author} {\bibinfo {author} {\bibfnamefont {W.}~\bibnamefont
  {Schirmacher}}, \bibinfo {author} {\bibfnamefont {B.}~\bibnamefont {Schmid}},
  \bibinfo {author} {\bibfnamefont {C.}~\bibnamefont {Tomaras}}, \bibinfo
  {author} {\bibfnamefont {G.}~\bibnamefont {Viliani}}, \bibinfo {author}
  {\bibfnamefont {G.}~\bibnamefont {Baldi}}, \bibinfo {author} {\bibfnamefont
  {G.}~\bibnamefont {Ruocco}}, \ and\ \bibinfo {author} {\bibfnamefont
  {T.}~\bibnamefont {Scopigno}},\ }\href {\doibase 10.1002/pssc.200777584}
  {\bibfield  {journal} {\bibinfo  {journal} {Phys. Status Solidi C}\ }\textbf
  {\bibinfo {volume} {5}},\ \bibinfo {pages} {862} (\bibinfo {year}
  {2008})}\BibitemShut {NoStop}%
\bibitem [{\citenamefont {Marruzzo}\ \emph {et~al.}(2013)\citenamefont
  {Marruzzo}, \citenamefont {Schirmacher}, \citenamefont {Fratalocchi},\ and\
  \citenamefont {Ruocco}}]{Marruzzo2013}%
  \BibitemOpen
  \bibfield  {author} {\bibinfo {author} {\bibfnamefont {A.}~\bibnamefont
  {Marruzzo}}, \bibinfo {author} {\bibfnamefont {W.}~\bibnamefont
  {Schirmacher}}, \bibinfo {author} {\bibfnamefont {A.}~\bibnamefont
  {Fratalocchi}}, \ and\ \bibinfo {author} {\bibfnamefont {G.}~\bibnamefont
  {Ruocco}},\ }\href {\doibase 10.1038/srep01407} {\bibfield  {journal}
  {\bibinfo  {journal} {Sci. Rep.}\ }\textbf {\bibinfo {volume} {3}},\ \bibinfo
  {pages} {1407} (\bibinfo {year} {2013})}\BibitemShut {NoStop}%
\bibitem [{\citenamefont {Schirmacher}\ \emph {et~al.}(2015)\citenamefont
  {Schirmacher}, \citenamefont {Scopigno},\ and\ \citenamefont
  {Ruocco}}]{Schirmacher2015}%
  \BibitemOpen
  \bibfield  {author} {\bibinfo {author} {\bibfnamefont {W.}~\bibnamefont
  {Schirmacher}}, \bibinfo {author} {\bibfnamefont {T.}~\bibnamefont
  {Scopigno}}, \ and\ \bibinfo {author} {\bibfnamefont {G.}~\bibnamefont
  {Ruocco}},\ }\href {\doibase 10.1016/j.jnoncrysol.2014.09.054} {\bibfield
  {journal} {\bibinfo  {journal} {J. Non-Cryst. Solids}\ }\textbf {\bibinfo
  {volume} {407}},\ \bibinfo {pages} {133} (\bibinfo {year}
  {2015})}\BibitemShut {NoStop}%
\bibitem [{\citenamefont {Tanguy}\ \emph {et~al.}(2010)\citenamefont {Tanguy},
  \citenamefont {Mantisi},\ and\ \citenamefont {Tsamados}}]{Tanguy2010}%
  \BibitemOpen
  \bibfield  {author} {\bibinfo {author} {\bibfnamefont {A.}~\bibnamefont
  {Tanguy}}, \bibinfo {author} {\bibfnamefont {B.}~\bibnamefont {Mantisi}}, \
  and\ \bibinfo {author} {\bibfnamefont {M.}~\bibnamefont {Tsamados}},\ }\href
  {\doibase 10.1209/0295-5075/90/16004} {\bibfield  {journal} {\bibinfo
  {journal} {EPL}\ }\textbf {\bibinfo {volume} {90}},\ \bibinfo {pages} {16004}
  (\bibinfo {year} {2010})}\BibitemShut {NoStop}%
\bibitem [{\citenamefont {Derlet}\ \emph {et~al.}(2012)\citenamefont {Derlet},
  \citenamefont {Maa\ss{}},\ and\ \citenamefont {L\"offler}}]{Derlet2012}%
  \BibitemOpen
  \bibfield  {author} {\bibinfo {author} {\bibfnamefont {P.~M.}\ \bibnamefont
  {Derlet}}, \bibinfo {author} {\bibfnamefont {R.}~\bibnamefont {Maa\ss{}}}, \
  and\ \bibinfo {author} {\bibfnamefont {J.~F.}\ \bibnamefont {L\"offler}},\
  }\href {\doibase 10.1140/epjb/e2012-20902-0} {\bibfield  {journal} {\bibinfo
  {journal} {EPJ B}\ }\textbf {\bibinfo {volume} {85}},\ \bibinfo {pages} {148}
  (\bibinfo {year} {2012})}\BibitemShut {NoStop}%
\bibitem [{\citenamefont {Ding}\ \emph {et~al.}(2014)\citenamefont {Ding},
  \citenamefont {Patinet}, \citenamefont {Falk}, \citenamefont {Cheng},\ and\
  \citenamefont {Ma}}]{Ding2014b}%
  \BibitemOpen
  \bibfield  {author} {\bibinfo {author} {\bibfnamefont {J.}~\bibnamefont
  {Ding}}, \bibinfo {author} {\bibfnamefont {S.}~\bibnamefont {Patinet}},
  \bibinfo {author} {\bibfnamefont {M.~L.}\ \bibnamefont {Falk}}, \bibinfo
  {author} {\bibfnamefont {Y.}~\bibnamefont {Cheng}}, \ and\ \bibinfo {author}
  {\bibfnamefont {E.}~\bibnamefont {Ma}},\ }\href {\doibase
  10.1073/pnas.1412095111} {\bibfield  {journal} {\bibinfo  {journal} {Proc.
  Natl. Acad. Sci. USA}\ }\textbf {\bibinfo {volume} {111}},\ \bibinfo {pages}
  {14052} (\bibinfo {year} {2014})}\BibitemShut {NoStop}%
\bibitem [{\citenamefont {L\'eonforte}\ \emph {et~al.}(2006)\citenamefont
  {L\'eonforte}, \citenamefont {Tanguy}, \citenamefont {Wittmer},\ and\
  \citenamefont {Barrat}}]{Leonforte2006}%
  \BibitemOpen
  \bibfield  {author} {\bibinfo {author} {\bibfnamefont {F.}~\bibnamefont
  {L\'eonforte}}, \bibinfo {author} {\bibfnamefont {A.}~\bibnamefont {Tanguy}},
  \bibinfo {author} {\bibfnamefont {J.~P.}\ \bibnamefont {Wittmer}}, \ and\
  \bibinfo {author} {\bibfnamefont {J.-L.}\ \bibnamefont {Barrat}},\ }\href
  {\doibase 10.1103/PhysRevLett.97.055501} {\bibfield  {journal} {\bibinfo
  {journal} {Phys. Rev. Lett.}\ }\textbf {\bibinfo {volume} {97}},\ \bibinfo
  {pages} {055501} (\bibinfo {year} {2006})}\BibitemShut {NoStop}%
\bibitem [{\citenamefont {Fusco}\ \emph {et~al.}(2010)\citenamefont {Fusco},
  \citenamefont {Albaret},\ and\ \citenamefont {Tanguy}}]{Fusco2010}%
  \BibitemOpen
  \bibfield  {author} {\bibinfo {author} {\bibfnamefont {C.}~\bibnamefont
  {Fusco}}, \bibinfo {author} {\bibfnamefont {T.}~\bibnamefont {Albaret}}, \
  and\ \bibinfo {author} {\bibfnamefont {A.}~\bibnamefont {Tanguy}},\ }\href
  {\doibase 10.1103/PhysRevE.82.066116} {\bibfield  {journal} {\bibinfo
  {journal} {Phys. Rev. E}\ }\textbf {\bibinfo {volume} {82}},\ \bibinfo
  {pages} {066116} (\bibinfo {year} {2010})}\BibitemShut {NoStop}%
\bibitem [{\citenamefont {Beltukov}\ \emph {et~al.}(2016)\citenamefont
  {Beltukov}, \citenamefont {Fusco}, \citenamefont {Parshin},\ and\
  \citenamefont {Tanguy}}]{Beltukov2016}%
  \BibitemOpen
  \bibfield  {author} {\bibinfo {author} {\bibfnamefont {Y.~M.}\ \bibnamefont
  {Beltukov}}, \bibinfo {author} {\bibfnamefont {C.}~\bibnamefont {Fusco}},
  \bibinfo {author} {\bibfnamefont {D.~A.}\ \bibnamefont {Parshin}}, \ and\
  \bibinfo {author} {\bibfnamefont {A.}~\bibnamefont {Tanguy}},\ }\href
  {\doibase 10.1103/PhysRevE.93.023006} {\bibfield  {journal} {\bibinfo
  {journal} {Phys. Rev. E}\ }\textbf {\bibinfo {volume} {93}},\ \bibinfo
  {pages} {023006} (\bibinfo {year} {2016})}\BibitemShut {NoStop}%
\bibitem [{\citenamefont {Taraskin}\ \emph {et~al.}(2001)\citenamefont
  {Taraskin}, \citenamefont {Loh}, \citenamefont {Natarajan},\ and\
  \citenamefont {Elliott}}]{Taraskin2001}%
  \BibitemOpen
  \bibfield  {author} {\bibinfo {author} {\bibfnamefont {S.~N.}\ \bibnamefont
  {Taraskin}}, \bibinfo {author} {\bibfnamefont {Y.~L.}\ \bibnamefont {Loh}},
  \bibinfo {author} {\bibfnamefont {G.}~\bibnamefont {Natarajan}}, \ and\
  \bibinfo {author} {\bibfnamefont {S.~R.}\ \bibnamefont {Elliott}},\ }\href
  {\doibase 10.1103/PhysRevLett.86.1255} {\bibfield  {journal} {\bibinfo
  {journal} {Phys. Rev. Lett.}\ }\textbf {\bibinfo {volume} {86}},\ \bibinfo
  {pages} {1255} (\bibinfo {year} {2001})}\BibitemShut {NoStop}%
\bibitem [{\citenamefont {Chumakov}\ \emph {et~al.}(2011)\citenamefont
  {Chumakov}, \citenamefont {Monaco}, \citenamefont {Monaco}, \citenamefont
  {Crichton}, \citenamefont {Bosak}, \citenamefont {R\"uffer}, \citenamefont
  {Meyer}, \citenamefont {Kargl}, \citenamefont {Comez}, \citenamefont
  {Fioretto}, \citenamefont {Giefers}, \citenamefont {Roitsch}, \citenamefont
  {Wortmann}, \citenamefont {Manghnani}, \citenamefont {Hushur}, \citenamefont
  {Williams}, \citenamefont {Balogh}, \citenamefont {Parli\'{n}ski},
  \citenamefont {Jochym},\ and\ \citenamefont {Piekarz}}]{Chumakov2011}%
  \BibitemOpen
  \bibfield  {author} {\bibinfo {author} {\bibfnamefont {A.~I.}\ \bibnamefont
  {Chumakov}}, \bibinfo {author} {\bibfnamefont {G.}~\bibnamefont {Monaco}},
  \bibinfo {author} {\bibfnamefont {A.}~\bibnamefont {Monaco}}, \bibinfo
  {author} {\bibfnamefont {W.~A.}\ \bibnamefont {Crichton}}, \bibinfo {author}
  {\bibfnamefont {A.}~\bibnamefont {Bosak}}, \bibinfo {author} {\bibfnamefont
  {R.}~\bibnamefont {R\"uffer}}, \bibinfo {author} {\bibfnamefont
  {A.}~\bibnamefont {Meyer}}, \bibinfo {author} {\bibfnamefont
  {F.}~\bibnamefont {Kargl}}, \bibinfo {author} {\bibfnamefont
  {L.}~\bibnamefont {Comez}}, \bibinfo {author} {\bibfnamefont
  {D.}~\bibnamefont {Fioretto}}, \bibinfo {author} {\bibfnamefont
  {H.}~\bibnamefont {Giefers}}, \bibinfo {author} {\bibfnamefont
  {S.}~\bibnamefont {Roitsch}}, \bibinfo {author} {\bibfnamefont
  {G.}~\bibnamefont {Wortmann}}, \bibinfo {author} {\bibfnamefont {M.~H.}\
  \bibnamefont {Manghnani}}, \bibinfo {author} {\bibfnamefont {A.}~\bibnamefont
  {Hushur}}, \bibinfo {author} {\bibfnamefont {Q.}~\bibnamefont {Williams}},
  \bibinfo {author} {\bibfnamefont {J.}~\bibnamefont {Balogh}}, \bibinfo
  {author} {\bibfnamefont {K.}~\bibnamefont {Parli\'{n}ski}}, \bibinfo {author}
  {\bibfnamefont {P.}~\bibnamefont {Jochym}}, \ and\ \bibinfo {author}
  {\bibfnamefont {P.}~\bibnamefont {Piekarz}},\ }\href {\doibase
  10.1103/PhysRevLett.106.225501} {\bibfield  {journal} {\bibinfo  {journal}
  {Phys. Rev. Lett.}\ }\textbf {\bibinfo {volume} {106}},\ \bibinfo {pages}
  {225501} (\bibinfo {year} {2011})}\BibitemShut {NoStop}%
\bibitem [{\citenamefont {Chumakov}\ \emph {et~al.}(2014)\citenamefont
  {Chumakov}, \citenamefont {Monaco}, \citenamefont {Fontana}, \citenamefont
  {Bosak}, \citenamefont {Hermann}, \citenamefont {Bessas}, \citenamefont
  {Wehinger}, \citenamefont {Crichton}, \citenamefont {Krisch}, \citenamefont
  {R\"uffer}, \citenamefont {Baldi}, \citenamefont {Carini~Jr.}, \citenamefont
  {Carini}, \citenamefont {D'Angelo}, \citenamefont {Gilioli}, \citenamefont
  {Tripodo}, \citenamefont {Zanatta}, \citenamefont {Winkler}, \citenamefont
  {Milman}, \citenamefont {Refson}, \citenamefont {Dove}, \citenamefont
  {Dubrovinskaia}, \citenamefont {Dubrovinsky}, \citenamefont {Keding},\ and\
  \citenamefont {Yue}}]{Chumakov2014}%
  \BibitemOpen
  \bibfield  {author} {\bibinfo {author} {\bibfnamefont {A.~I.}\ \bibnamefont
  {Chumakov}}, \bibinfo {author} {\bibfnamefont {G.}~\bibnamefont {Monaco}},
  \bibinfo {author} {\bibfnamefont {A.}~\bibnamefont {Fontana}}, \bibinfo
  {author} {\bibfnamefont {A.}~\bibnamefont {Bosak}}, \bibinfo {author}
  {\bibfnamefont {R.~P.}\ \bibnamefont {Hermann}}, \bibinfo {author}
  {\bibfnamefont {D.}~\bibnamefont {Bessas}}, \bibinfo {author} {\bibfnamefont
  {B.}~\bibnamefont {Wehinger}}, \bibinfo {author} {\bibfnamefont {W.~A.}\
  \bibnamefont {Crichton}}, \bibinfo {author} {\bibfnamefont {M.}~\bibnamefont
  {Krisch}}, \bibinfo {author} {\bibfnamefont {R.}~\bibnamefont {R\"uffer}},
  \bibinfo {author} {\bibfnamefont {G.}~\bibnamefont {Baldi}}, \bibinfo
  {author} {\bibfnamefont {G.}~\bibnamefont {Carini~Jr.}}, \bibinfo {author}
  {\bibfnamefont {G.}~\bibnamefont {Carini}}, \bibinfo {author} {\bibfnamefont
  {G.}~\bibnamefont {D'Angelo}}, \bibinfo {author} {\bibfnamefont
  {E.}~\bibnamefont {Gilioli}}, \bibinfo {author} {\bibfnamefont
  {G.}~\bibnamefont {Tripodo}}, \bibinfo {author} {\bibfnamefont
  {M.}~\bibnamefont {Zanatta}}, \bibinfo {author} {\bibfnamefont
  {B.}~\bibnamefont {Winkler}}, \bibinfo {author} {\bibfnamefont
  {V.}~\bibnamefont {Milman}}, \bibinfo {author} {\bibfnamefont
  {K.}~\bibnamefont {Refson}}, \bibinfo {author} {\bibfnamefont {M.~T.}\
  \bibnamefont {Dove}}, \bibinfo {author} {\bibfnamefont {N.}~\bibnamefont
  {Dubrovinskaia}}, \bibinfo {author} {\bibfnamefont {L.}~\bibnamefont
  {Dubrovinsky}}, \bibinfo {author} {\bibfnamefont {R.}~\bibnamefont {Keding}},
  \ and\ \bibinfo {author} {\bibfnamefont {Y.~Z.}\ \bibnamefont {Yue}},\ }\href
  {\doibase 10.1103/PhysRevLett.112.025502} {\bibfield  {journal} {\bibinfo
  {journal} {Phys. Rev. Lett.}\ }\textbf {\bibinfo {volume} {112}},\ \bibinfo
  {pages} {025502} (\bibinfo {year} {2014})}\BibitemShut {NoStop}%
\bibitem [{\citenamefont {Chumakov}\ and\ \citenamefont
  {Monaco}(2015)}]{Chumakov2015}%
  \BibitemOpen
  \bibfield  {author} {\bibinfo {author} {\bibfnamefont {A.~I.}\ \bibnamefont
  {Chumakov}}\ and\ \bibinfo {author} {\bibfnamefont {G.}~\bibnamefont
  {Monaco}},\ }\href {\doibase 10.1016/j.jnoncrysol.2014.09.031} {\bibfield
  {journal} {\bibinfo  {journal} {J. Non-Cryst. Solids}\ }\textbf {\bibinfo
  {volume} {407}},\ \bibinfo {pages} {126} (\bibinfo {year}
  {2015})}\BibitemShut {NoStop}%
\bibitem [{\citenamefont {Chumakov}\ \emph {et~al.}(2016)\citenamefont
  {Chumakov}, \citenamefont {Monaco}, \citenamefont {Han}, \citenamefont {Xi},
  \citenamefont {Bosak}, \citenamefont {Paolasini}, \citenamefont
  {Chernyshov},\ and\ \citenamefont {Dyadkin}}]{Chumakov2016}%
  \BibitemOpen
  \bibfield  {author} {\bibinfo {author} {\bibfnamefont {A.~I.}\ \bibnamefont
  {Chumakov}}, \bibinfo {author} {\bibfnamefont {G.}~\bibnamefont {Monaco}},
  \bibinfo {author} {\bibfnamefont {X.}~\bibnamefont {Han}}, \bibinfo {author}
  {\bibfnamefont {L.}~\bibnamefont {Xi}}, \bibinfo {author} {\bibfnamefont
  {A.}~\bibnamefont {Bosak}}, \bibinfo {author} {\bibfnamefont
  {L.}~\bibnamefont {Paolasini}}, \bibinfo {author} {\bibfnamefont
  {D.}~\bibnamefont {Chernyshov}}, \ and\ \bibinfo {author} {\bibfnamefont
  {V.}~\bibnamefont {Dyadkin}},\ }\href {\doibase
  10.1080/14786435.2015.1108528} {\bibfield  {journal} {\bibinfo  {journal}
  {Philos. Mag.}\ }\textbf {\bibinfo {volume} {96}},\ \bibinfo {pages} {743}
  (\bibinfo {year} {2016})}\BibitemShut {NoStop}%
\bibitem [{\citenamefont {Baldi}\ \emph {et~al.}(2016)\citenamefont {Baldi},
  \citenamefont {Carini~Jr.}, \citenamefont {Carini}, \citenamefont {Chumakov},
  \citenamefont {Maschio}, \citenamefont {D'Angelo}, \citenamefont {Fontana},
  \citenamefont {Gilioli}, \citenamefont {Monaco}, \citenamefont {Orsingher},
  \citenamefont {Rossi},\ and\ \citenamefont {Zanatta}}]{Baldi2016}%
  \BibitemOpen
  \bibfield  {author} {\bibinfo {author} {\bibfnamefont {G.}~\bibnamefont
  {Baldi}}, \bibinfo {author} {\bibfnamefont {G.}~\bibnamefont {Carini~Jr.}},
  \bibinfo {author} {\bibfnamefont {G.}~\bibnamefont {Carini}}, \bibinfo
  {author} {\bibfnamefont {A.}~\bibnamefont {Chumakov}}, \bibinfo {author}
  {\bibfnamefont {R.~D.}\ \bibnamefont {Maschio}}, \bibinfo {author}
  {\bibfnamefont {G.}~\bibnamefont {D'Angelo}}, \bibinfo {author}
  {\bibfnamefont {A.}~\bibnamefont {Fontana}}, \bibinfo {author} {\bibfnamefont
  {E.}~\bibnamefont {Gilioli}}, \bibinfo {author} {\bibfnamefont
  {G.}~\bibnamefont {Monaco}}, \bibinfo {author} {\bibfnamefont
  {L.}~\bibnamefont {Orsingher}}, \bibinfo {author} {\bibfnamefont
  {B.}~\bibnamefont {Rossi}}, \ and\ \bibinfo {author} {\bibfnamefont
  {M.}~\bibnamefont {Zanatta}},\ }\href {\doibase
  10.1080/14786435.2015.1127445} {\bibfield  {journal} {\bibinfo  {journal}
  {Philos. Mag.}\ }\textbf {\bibinfo {volume} {96}},\ \bibinfo {pages} {754}
  (\bibinfo {year} {2016})}\BibitemShut {NoStop}%
\bibitem [{\citenamefont {Yeh}(2013)}]{Yeh2013}%
  \BibitemOpen
  \bibfield  {author} {\bibinfo {author} {\bibfnamefont {J.-W.}\ \bibnamefont
  {Yeh}},\ }\href {\doibase 10.1007/s11837-013-0761-6} {\bibfield  {journal}
  {\bibinfo  {journal} {JOM}\ }\textbf {\bibinfo {volume} {65}},\ \bibinfo
  {pages} {1759} (\bibinfo {year} {2013})}\BibitemShut {NoStop}%
\bibitem [{\citenamefont {Tsai}\ and\ \citenamefont {Yeh}(2014)}]{Tsai2014}%
  \BibitemOpen
  \bibfield  {author} {\bibinfo {author} {\bibfnamefont {M.-H.}\ \bibnamefont
  {Tsai}}\ and\ \bibinfo {author} {\bibfnamefont {J.-W.}\ \bibnamefont {Yeh}},\
  }\href {\doibase 10.1080/21663831.2014.912690} {\bibfield  {journal}
  {\bibinfo  {journal} {Mater. Res. Lett.}\ }\textbf {\bibinfo {volume} {2}},\
  \bibinfo {pages} {107} (\bibinfo {year} {2014})}\BibitemShut {NoStop}%
\bibitem [{\citenamefont {Miracle}\ \emph {et~al.}(2014)\citenamefont
  {Miracle}, \citenamefont {Miller}, \citenamefont {Senkov}, \citenamefont
  {Woodward}, \citenamefont {Uchic},\ and\ \citenamefont
  {Tiley}}]{Miracle2014}%
  \BibitemOpen
  \bibfield  {author} {\bibinfo {author} {\bibfnamefont {D.~B.}\ \bibnamefont
  {Miracle}}, \bibinfo {author} {\bibfnamefont {J.~D.}\ \bibnamefont {Miller}},
  \bibinfo {author} {\bibfnamefont {O.~N.}\ \bibnamefont {Senkov}}, \bibinfo
  {author} {\bibfnamefont {C.}~\bibnamefont {Woodward}}, \bibinfo {author}
  {\bibfnamefont {M.~D.}\ \bibnamefont {Uchic}}, \ and\ \bibinfo {author}
  {\bibfnamefont {J.}~\bibnamefont {Tiley}},\ }\href {\doibase
  10.3390/e16010494} {\bibfield  {journal} {\bibinfo  {journal} {Entropy}\
  }\textbf {\bibinfo {volume} {16}},\ \bibinfo {pages} {494} (\bibinfo {year}
  {2014})}\BibitemShut {NoStop}%
\bibitem [{\citenamefont {Khonik}(2015)}]{Khonik2015}%
  \BibitemOpen
  \bibfield  {author} {\bibinfo {author} {\bibfnamefont {V.~A.}\ \bibnamefont
  {Khonik}},\ }\href {\doibase 10.3390/met5020504} {\bibfield  {journal}
  {\bibinfo  {journal} {Metals}\ }\textbf {\bibinfo {volume} {5}},\ \bibinfo
  {pages} {504} (\bibinfo {year} {2015})}\BibitemShut {NoStop}%
\bibitem [{\citenamefont {Zhou}\ \emph {et~al.}(2004)\citenamefont {Zhou},
  \citenamefont {Johnson},\ and\ \citenamefont {Wadley}}]{Zhou2004}%
  \BibitemOpen
  \bibfield  {author} {\bibinfo {author} {\bibfnamefont {X.~W.}\ \bibnamefont
  {Zhou}}, \bibinfo {author} {\bibfnamefont {R.~A.}\ \bibnamefont {Johnson}}, \
  and\ \bibinfo {author} {\bibfnamefont {H.~N.~G.}\ \bibnamefont {Wadley}},\
  }\href {\doibase 10.1103/PhysRevB.69.144113} {\bibfield  {journal} {\bibinfo
  {journal} {Phys. Rev. B}\ }\textbf {\bibinfo {volume} {69}},\ \bibinfo
  {pages} {144113} (\bibinfo {year} {2004})}\BibitemShut {NoStop}%
\bibitem [{\citenamefont {Plimpton}(1995)}]{Plimpton1995}%
  \BibitemOpen
  \bibfield  {author} {\bibinfo {author} {\bibfnamefont {S.}~\bibnamefont
  {Plimpton}},\ }\href {\doibase 10.1006/jcph.1995.1039} {\bibfield  {journal}
  {\bibinfo  {journal} {J. Comp. Phys.}\ }\textbf {\bibinfo {volume} {117}},\
  \bibinfo {pages} {1} (\bibinfo {year} {1995})},\ \bibinfo {note}
  {\url{http://lammps.sandia.gov/}}\BibitemShut {NoStop}%
\bibitem [{\citenamefont {Sadigh}\ \emph {et~al.}(2012)\citenamefont {Sadigh},
  \citenamefont {Erhart}, \citenamefont {Stukowski}, \citenamefont {Caro},
  \citenamefont {Martinez},\ and\ \citenamefont {Zepeda-Ruiz}}]{Sadigh2012}%
  \BibitemOpen
  \bibfield  {author} {\bibinfo {author} {\bibfnamefont {B.}~\bibnamefont
  {Sadigh}}, \bibinfo {author} {\bibfnamefont {P.}~\bibnamefont {Erhart}},
  \bibinfo {author} {\bibfnamefont {A.}~\bibnamefont {Stukowski}}, \bibinfo
  {author} {\bibfnamefont {A.}~\bibnamefont {Caro}}, \bibinfo {author}
  {\bibfnamefont {E.}~\bibnamefont {Martinez}}, \ and\ \bibinfo {author}
  {\bibfnamefont {L.}~\bibnamefont {Zepeda-Ruiz}},\ }\href {\doibase
  10.1103/PhysRevB.85.184203} {\bibfield  {journal} {\bibinfo  {journal} {Phys.
  Rev. B}\ }\textbf {\bibinfo {volume} {85}},\ \bibinfo {pages} {184203}
  (\bibinfo {year} {2012})}\BibitemShut {NoStop}%
\bibitem [{\citenamefont {Wilson}(2014)}]{Wilson2014}%
  \BibitemOpen
  \bibfield  {author} {\bibinfo {author} {\bibfnamefont {A.~G.}\ \bibnamefont
  {Wilson}},\ }\emph {\bibinfo {title} {Covariance kernels for fast automatic
  pattern discovery and extrapolation with Gaussian processes}},\ \href@noop {}
  {Ph.D. thesis},\ \bibinfo  {school} {Trinity College, University of
  Cambridge} (\bibinfo {year} {2014})\BibitemShut {NoStop}%
\bibitem [{\citenamefont {Dickey}\ and\ \citenamefont
  {Paskin}(1969)}]{Dickey1969}%
  \BibitemOpen
  \bibfield  {author} {\bibinfo {author} {\bibfnamefont {J.~M.}\ \bibnamefont
  {Dickey}}\ and\ \bibinfo {author} {\bibfnamefont {A.}~\bibnamefont
  {Paskin}},\ }\href {\doibase 10.1103/PhysRev.188.1407} {\bibfield  {journal}
  {\bibinfo  {journal} {Phys. Rev.}\ }\textbf {\bibinfo {volume} {188}},\
  \bibinfo {pages} {1407} (\bibinfo {year} {1969})}\BibitemShut {NoStop}%
\bibitem [{\citenamefont {Pathria}(1996)}]{Pathria1996}%
  \BibitemOpen
  \bibfield  {author} {\bibinfo {author} {\bibfnamefont {R.~K.}\ \bibnamefont
  {Pathria}},\ }\href@noop {} {\emph {\bibinfo {title} {Statistical
  Mechanics}}},\ \bibinfo {edition} {2nd}\ ed.\ (\bibinfo  {publisher}
  {Elsevier Butterworth-Heinemann},\ \bibinfo {address} {Oxford},\ \bibinfo
  {year} {1996})\BibitemShut {NoStop}%
\bibitem [{\citenamefont {Shimizu}\ \emph {et~al.}(2007)\citenamefont
  {Shimizu}, \citenamefont {Ogata},\ and\ \citenamefont {Li}}]{Shimizu2007}%
  \BibitemOpen
  \bibfield  {author} {\bibinfo {author} {\bibfnamefont {F.}~\bibnamefont
  {Shimizu}}, \bibinfo {author} {\bibfnamefont {S.}~\bibnamefont {Ogata}}, \
  and\ \bibinfo {author} {\bibfnamefont {J.}~\bibnamefont {Li}},\ }\href
  {\doibase 10.2320/matertrans.MJ200769} {\bibfield  {journal} {\bibinfo
  {journal} {Mater. Trans.}\ }\textbf {\bibinfo {volume} {48}},\ \bibinfo
  {pages} {2923} (\bibinfo {year} {2007})}\BibitemShut {NoStop}%
\bibitem [{\citenamefont {Stukowski}(2010)}]{Stukowski2010}%
  \BibitemOpen
  \bibfield  {author} {\bibinfo {author} {\bibfnamefont {A.}~\bibnamefont
  {Stukowski}},\ }\href {\doibase 10.1088/0965-0393/18/1/015012} {\bibfield
  {journal} {\bibinfo  {journal} {Model. Simul. Mater. Sci. Eng.}\ }\textbf
  {\bibinfo {volume} {18}},\ \bibinfo {pages} {015012} (\bibinfo {year}
  {2010})},\ \bibinfo {note} {\url{http://ovito.org/}}\BibitemShut {NoStop}%
\bibitem [{\citenamefont {Thomson}(1856)}]{Thomson1856}%
  \BibitemOpen
  \bibfield  {author} {\bibinfo {author} {\bibfnamefont {W.}~\bibnamefont
  {Thomson}},\ }\href {\doibase 10.1098/rstl.1856.0022} {\bibfield  {journal}
  {\bibinfo  {journal} {Philos. Trans. Roy. Soc. London}\ }\textbf {\bibinfo
  {volume} {146}},\ \bibinfo {pages} {481} (\bibinfo {year}
  {1856})}\BibitemShut {NoStop}%
\bibitem [{\citenamefont {Honeycutt}\ and\ \citenamefont
  {Andersen}(1987)}]{Honeycutt1987}%
  \BibitemOpen
  \bibfield  {author} {\bibinfo {author} {\bibfnamefont {J.~D.}\ \bibnamefont
  {Honeycutt}}\ and\ \bibinfo {author} {\bibfnamefont {H.~C.}\ \bibnamefont
  {Andersen}},\ }\href {\doibase 10.1021/j100303a014} {\bibfield  {journal}
  {\bibinfo  {journal} {J. Phys. Chem.}\ }\textbf {\bibinfo {volume} {91}},\
  \bibinfo {pages} {4950} (\bibinfo {year} {1987})}\BibitemShut {NoStop}%
\bibitem [{\citenamefont {Stukowski}(2012)}]{Stukowski2012}%
  \BibitemOpen
  \bibfield  {author} {\bibinfo {author} {\bibfnamefont {A.}~\bibnamefont
  {Stukowski}},\ }\href {\doibase 10.1088/0965-0393/20/4/045021} {\bibfield
  {journal} {\bibinfo  {journal} {Model. Simul. Mater. Sc.}\ }\textbf {\bibinfo
  {volume} {20}},\ \bibinfo {pages} {045021} (\bibinfo {year}
  {2012})}\BibitemShut {NoStop}%
\bibitem [{\citenamefont {Vorono\"i}(1908{\natexlab{a}})}]{Voronoi1908}%
  \BibitemOpen
  \bibfield  {author} {\bibinfo {author} {\bibfnamefont {G.}~\bibnamefont
  {Vorono\"i}},\ }\href {\doibase 10.1515/crll.1908.133.97} {\bibfield
  {journal} {\bibinfo  {journal} {J. Reine Angew. Math.}\ }\textbf {\bibinfo
  {volume} {133}},\ \bibinfo {pages} {97} (\bibinfo {year}
  {1908}{\natexlab{a}})}\BibitemShut {NoStop}%
\bibitem [{\citenamefont {Vorono\"i}(1908{\natexlab{b}})}]{Voronoi1908a}%
  \BibitemOpen
  \bibfield  {author} {\bibinfo {author} {\bibfnamefont {G.}~\bibnamefont
  {Vorono\"i}},\ }\href {\doibase 10.1515/crll.1908.134.198} {\bibfield
  {journal} {\bibinfo  {journal} {J. Reine Angew. Math.}\ }\textbf {\bibinfo
  {volume} {134}},\ \bibinfo {pages} {198} (\bibinfo {year}
  {1908}{\natexlab{b}})}\BibitemShut {NoStop}%
\bibitem [{\citenamefont {Vorono\"i}(1909)}]{Voronoi1909}%
  \BibitemOpen
  \bibfield  {author} {\bibinfo {author} {\bibfnamefont {G.}~\bibnamefont
  {Vorono\"i}},\ }\href {\doibase 10.1515/crll.1909.136.67} {\bibfield
  {journal} {\bibinfo  {journal} {J. Reine Angew. Math.}\ }\textbf {\bibinfo
  {volume} {136}},\ \bibinfo {pages} {67} (\bibinfo {year} {1909})}\BibitemShut
  {NoStop}%
\bibitem [{\citenamefont {Brostow}\ \emph {et~al.}(1998)\citenamefont
  {Brostow}, \citenamefont {Chybicki}, \citenamefont {Laskowski},\ and\
  \citenamefont {Rybicki}}]{Brostow1998}%
  \BibitemOpen
  \bibfield  {author} {\bibinfo {author} {\bibfnamefont {W.}~\bibnamefont
  {Brostow}}, \bibinfo {author} {\bibfnamefont {M.}~\bibnamefont {Chybicki}},
  \bibinfo {author} {\bibfnamefont {R.}~\bibnamefont {Laskowski}}, \ and\
  \bibinfo {author} {\bibfnamefont {J.}~\bibnamefont {Rybicki}},\ }\href
  {\doibase 10.1103/PhysRevB.57.13448} {\bibfield  {journal} {\bibinfo
  {journal} {Phys. Rev. B}\ }\textbf {\bibinfo {volume} {57}},\ \bibinfo
  {pages} {13448} (\bibinfo {year} {1998})}\BibitemShut {NoStop}%
\bibitem [{\citenamefont {Inoue}(2000)}]{Inoue2000}%
  \BibitemOpen
  \bibfield  {author} {\bibinfo {author} {\bibfnamefont {A.}~\bibnamefont
  {Inoue}},\ }\href {\doibase 10.1016/S1359-6454(99)00300-6} {\bibfield
  {journal} {\bibinfo  {journal} {Acta Mater.}\ }\textbf {\bibinfo {volume}
  {48}},\ \bibinfo {pages} {279} (\bibinfo {year} {2000})}\BibitemShut
  {NoStop}%
\bibitem [{\citenamefont {B\"unz}\ \emph {et~al.}(2014)\citenamefont {B\"unz},
  \citenamefont {Brink}, \citenamefont {Tsuchiya}, \citenamefont {Meng},
  \citenamefont {Wilde},\ and\ \citenamefont {Albe}}]{Bunz2014}%
  \BibitemOpen
  \bibfield  {author} {\bibinfo {author} {\bibfnamefont {J.}~\bibnamefont
  {B\"unz}}, \bibinfo {author} {\bibfnamefont {T.}~\bibnamefont {Brink}},
  \bibinfo {author} {\bibfnamefont {K.}~\bibnamefont {Tsuchiya}}, \bibinfo
  {author} {\bibfnamefont {F.}~\bibnamefont {Meng}}, \bibinfo {author}
  {\bibfnamefont {G.}~\bibnamefont {Wilde}}, \ and\ \bibinfo {author}
  {\bibfnamefont {K.}~\bibnamefont {Albe}},\ }\href {\doibase
  10.1103/PhysRevLett.112.135501} {\bibfield  {journal} {\bibinfo  {journal}
  {Phys. Rev. Lett.}\ }\textbf {\bibinfo {volume} {112}},\ \bibinfo {pages}
  {135501} (\bibinfo {year} {2014})}\BibitemShut {NoStop}%
\bibitem [{\citenamefont {Mitrofanov}\ \emph {et~al.}(2014)\citenamefont
  {Mitrofanov}, \citenamefont {Peterlechner}, \citenamefont {Divinski},\ and\
  \citenamefont {Wilde}}]{Mitrofanov2014}%
  \BibitemOpen
  \bibfield  {author} {\bibinfo {author} {\bibfnamefont {Y.~P.}\ \bibnamefont
  {Mitrofanov}}, \bibinfo {author} {\bibfnamefont {M.}~\bibnamefont
  {Peterlechner}}, \bibinfo {author} {\bibfnamefont {S.~V.}\ \bibnamefont
  {Divinski}}, \ and\ \bibinfo {author} {\bibfnamefont {G.}~\bibnamefont
  {Wilde}},\ }\href {\doibase 10.1103/PhysRevLett.112.135901} {\bibfield
  {journal} {\bibinfo  {journal} {Phys. Rev. Lett.}\ }\textbf {\bibinfo
  {volume} {112}},\ \bibinfo {pages} {135901} (\bibinfo {year}
  {2014})}\BibitemShut {NoStop}%
\bibitem [{\citenamefont {Ritter}\ and\ \citenamefont
  {Albe}(2011)}]{Ritter2011}%
  \BibitemOpen
  \bibfield  {author} {\bibinfo {author} {\bibfnamefont {Y.}~\bibnamefont
  {Ritter}}\ and\ \bibinfo {author} {\bibfnamefont {K.}~\bibnamefont {Albe}},\
  }\href {\doibase 10.1016/j.actamat.2011.07.063} {\bibfield  {journal}
  {\bibinfo  {journal} {Acta Mater.}\ }\textbf {\bibinfo {volume} {59}},\
  \bibinfo {pages} {7082} (\bibinfo {year} {2011})}\BibitemShut {NoStop}%
\bibitem [{\citenamefont {Ritter}\ and\ \citenamefont
  {Albe}(2012)}]{Ritter2012}%
  \BibitemOpen
  \bibfield  {author} {\bibinfo {author} {\bibfnamefont {Y.}~\bibnamefont
  {Ritter}}\ and\ \bibinfo {author} {\bibfnamefont {K.}~\bibnamefont {Albe}},\
  }\href {\doibase 10.1063/1.4717748} {\bibfield  {journal} {\bibinfo
  {journal} {J. Appl. Phys.}\ }\textbf {\bibinfo {volume} {111}},\ \bibinfo
  {pages} {103527} (\bibinfo {year} {2012})}\BibitemShut {NoStop}%
\end{thebibliography}
%

\end{document}